\documentclass[journal=jacsat,manuscript=article]{achemso}

\usepackage[version=4]{mhchem} % Formula subscripts using \ce{}
\usepackage{siunitx}
\DeclareSIUnit{\molecule}{molecule}
\DeclareSIUnit{\cal}{cal}
\usepackage{booktabs}
\usepackage{graphicx}
\usepackage{caption}
\usepackage{subcaption}
\usepackage{longtable}
\usepackage{lscape}
\usepackage{blindtext}

\author{Vincenzo Barone}
\email{vincenzo.barone@sns.it}
\author{Jacopo Lupi}
\author{Zoi Salta}
\author{Nicola Tasinato}
\affiliation
{SMART Laboratory, Scuola Normale Superiore di Pisa, piazza dei Cavalieri 7, 56125 Pisa, Italy}

\title[An \textsf{achemso} demo]
  {Development and validation of a parameter-free model chemistry for the computation of reliable reaction rates}

\abbreviations{IR,NMR,UV}

\begin{document}

%%%%%%%%%%%%%%%%%%%%%%%%%%%%%%%%%%%%%%%%%%%%%%%%%%%%%%%%%%%%%%%%%%%%%
%% The "tocentry" environment can be used to create an entry for the
%% graphical table of contents. It is given here as some journals
%% require that it is printed as part of the abstract page. It will
%% be automatically moved as appropriate.
%%%%%%%%%%%%%%%%%%%%%%%%%%%%%%%%%%%%%%%%%%%%%%%%%%%%%%%%%%%%%%%%%%%%%
\begin{tocentry}
\includegraphics{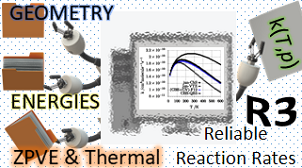}
\end{tocentry}

%%%%%%%%%%%%%%%%%%%%%%%%%%%%%%%%%%%%%%%%%%%%%%%%%%%%%%%%%%%%%%%%%%%%%
%% The abstract environment will automatically gobble the contents
%% if an abstract is not used by the target journal.
%%%%%%%%%%%%%%%%%%%%%%%%%%%%%%%%%%%%%%%%%%%%%%%%%%%%%%%%%%%%%%%%%%%%%
\begin{abstract}
 A recently developed model chemistry (jun-Cheap) has been slightly modified and proposed as an effective, reliable and parameter-free scheme for the computation of accurate reaction rates with special reference to astrochemical and atmospheric processes. Benchmarks with different sets of state-of-the-art energy barriers spanning a wide range of values show that, in the absence of strong multi-reference contributions, the proposed model outperforms the most well-known model chemistries, reaching a sub-chemical accuracy without any empirical parameter and with affordable  computer times. Some test cases show that geometries, energy barriers, zero point energies and thermal contributions computed at this level can be used in the framework of the master equation approach based on ab-initio transition state theory (AITSTME) for obtaining accurate reaction rates.
\end{abstract}

%%%%%%%%%%%%%%%%%%%%%%%%%%%%%%%%%%%%%%%%%%%%%%%%%%%%%%%%%%%%%%%%%%%%%
%% Start the main part of the manuscript here.
%%%%%%%%%%%%%%%%%%%%%%%%%%%%%%%%%%%%%%%%%%%%%%%%%%%%%%%%%%%%%%%%%%%%%
\section{Introduction}

For many years, scientists were skeptical about the presence of molecular systems in the interstellar space due to the harsh physical conditions (low temperature and pressure in the presence of high-energy radiation fields) characterizing this environment. However, contrary to those expectations, more than 200 molecules have been now identified in the interstellar and circumstellar medium (ISM),\cite{astrochymist} including several so-called complex organic interstellar molecules (iCOMs), namely molecules containing carbon and a total of more than 6 atoms.\cite{COMs} Most of the observed species should have a very short life-time according to earth-based  standards, but the inter-molecular processes leading to thermodynamic equilibrium are not effective in the ISM due to its extreme physical parameters.\cite{snow2006,yamamoto2017} This situation calls for a strong interplay between observations, laboratory studies and computational approaches to understand the chemical evolution in these regions and to explain the observed abundances of different species.

Astrochemical models are virtual laboratories including thousands of reactions and whose main goal is to reproduce the observational data to the best possible extent. Although the available astrochemical models show widely different degrees of sophistication\cite{DOBRIJEVIC2016313}, all of them share the same basic ingredients:\cite{carbo1985} a set of initial conditions (total density, temperature, etc.) and a panel of chemical reactions characterized by their respective temperature-dependent rate constants and most likely exit channels. In order to improve the current predictions provided by these models, the reactions responsible for the largest uncertainties on the abundances must be studied in more detail by laboratory experiments and/or theoretical methods to provide improved rate constants and branching ratios.

Chemical kinetics plays a fundamental role also in the different, but related context of  atmospheric models that 
try to reproduce and interpret the large number of chemical processes occurring in the troposphere.
Reaction rate coefficients and product yields have been traditionally obtained either by means of suitable experimental techniques\cite{atkinson2006} or estimated using structure-activity relationships.\cite{kwok1995} The massive number of organic compounds released in the atmosphere and the corresponding huge number of possible reactions ruling their oxidation/degradation pathways make experimental measurement of even a small fraction of key processes a daunting task.  In recent years, computational chemistry has begun to contribute substantially to a better understanding of several important reaction sequences in the atmosphere.\cite{vereecken2015} These contributions have, at their heart, the use of electronic structure calculations to determine the energies and other characteristics (mainly geometries and vibrational frequencies) of stable species, reactive complexes and transition states, which are then used in theoretical frameworks to determine rate coefficients. The main factor limiting the accuracy of this process is the computation of accurate values for all the energy barriers ruling the different elementary steps. Next, zero point energies and finite temperature contributions come into play, whose contributions can become non-negligible already for medium size systems. 

Several non-empirical procedures have been developed and employed for the generation of accurate thermochemical data, which for small systems come close to the full configuration interaction (FCI) complete basis set (CBS) limit.\cite{karton_wires} Among the most successful approaches there are the Weizmann-n series (with the most accurate being W4\cite{karton2012}), the focal point analysis (FPA) approach,\cite{laane2012structures,csaszar1998} the Feller-Dixon-Peterson model (FDP) \cite{fdp} and the extrapolated ab-initio thermochemistry (HEAT) protocol\cite{tajti2004,bomble2006,harding2008}.  A simplified version of the HEAT protocol is obtained by retaining only the extrapolation to the CBS limit at the CCSD(T) level and the incorporation of the core-valence corrections, thus leading to the model referred to in the following as CBS-CV. This approach is rather well tested in the literature and was shown to provide results with an accuracy well within \SI{0.5}{\kilo\cal\per\mol}. Recently, alternative protocols have been proposed, which employ explicitly-correlated approaches:\cite{dbh24_f12,karton_wires} thanks to the faster convergence to the complete basis set limit, these approaches allow some computer time saving, but the rate determining step remains the evaluation of higher level contributions. 

For larger molecular systems, more approximate composite methods are unavoidable, which aim at reaching the so-called chemical accuracy (\SI{1}{\kilo\cal\per\mol}). The most well known among these so-called model chemistries are the last versions of the Gn \cite{gn_wires} (G4 \cite{G4}) and CBS \cite{cbs94}(CBS-QB3 \cite{montgomery2000}) families. However, all these models include some empirical parameters and employ geometries, which are not fully reliable for transition states and non-covalent complexes ruling the entrance channels of most reactions of astrochemical and atmospheric interest. As a matter of fact, the most reliable protocols (e.g. HEAT) push geometry optimizations to the limit in order to obtain accurate energetics, whereas, at the other extreme, Gn and CBS-x schemes employ B3LYP geometries, whose accuracy is often unsatisfactory.\cite{curtiss10}

In the last few years, a reliable and accurate computational protocol, referred to as cheap scheme (ChS) and devoid of any empirical parameter, has been developed and tested with remarkable success for structural and energetic data.\cite{puzzarini2011, puzzarini2013,puzzarini2014}. In conjunction with geometries and harmonic frequencies issuing from double hybrid functionals, ChS has given promising results also for the activation energies of some reactions of astrochemical interest.\cite{Salta2020,lupi2020,apjl2020,baiano2020,ballotta2021} More recently, an improved variant (referred to as jun-Cheap scheme, jChS) has been introduced, which, thanks to the use of the 'june' partially augmented basis set of the 'calendar' family \cite{SummerBasisSets}, provides very accurate results also for non-covalent interactions.\cite {alessandrini2020,melli2021}. On these grounds, in this paper we provide a comprehensive benchmark of the jChS model chemistry for several classes of reactions for which accurate reference results are available or have been purposely computed. Together with electronic energies, we analyze also zero point energies, thermal contributions to enthalpies and entropies and overall reaction rates computed for elementary reactions in the framework of the master equation (ME) approach based on ab initio transition state theory (AITSTME) \cite{VTS_Truhlar,miller2006master,TS_klippenstein}.

The paper is organized as follows. In the first part we validate the jChS model chemistry with reference to some well-known databases, namely i) the 24 energy barriers available in the latest updated version of the DBH24 database,\cite{zheng2009} ii) the 52 barriers of Truhlar’s HTBH38\cite{zhao2005a} and NHTBH38\cite{zhao2005b} databases not included in DBH24 and iii) 7 representative reactions from Karton’s BH28 database.\cite{karton2019} 
When needed, the reference values are updated by new computations performed with a composite method closely resembling the W3.2 model \cite{karton2006}. 

Next, the reliability of the jChS model chemistry for zero point energies and thermal contributions to enthalpies and entropies is assessed with respect to  the new databases THCS21 and THOS10 containing accurate reference values for closed- and open-shell systems, respectively. 

Finally, the role of different contributions in determining the overall accuracy of computed reaction rates is analyzed by means of some simple elementary reactions and two more complex reaction networks relevant for astrochemistry and atmospheric chemistry. Conclusions and perspectives are given in the last section. 

\section{Computational details}
All the composite schemes employed in the present work extrapolate single point energies computed at suitable geometries (see next sections) using the cc-pV(n+$d$)Z (hereafter nZ) \cite{Dunning-JCP1989_cc-pVxZ} or
jun-cc-pV(n+$d$)Z (hereafter jnZ) \cite{SummerBasisSets} families of basis sets.
The coupled cluster (CC) ansatz including single, double and (perturbative) triple excitations (CCSD(T)) \cite{raghavachari1989fifth} within the frozen-core approximation and in conjunction with 3Z or j3Z basis sets is always employed in the first step. Next, CBS extrapolation and core-valence correlation (CV) are added using either MP2 \cite{moller1934note} (leading, in conjunction with jnZ basis sets, to our standard jChS model) or CCSD(T). In the latter case, inclusion of higher-level terms (diagonal Born-Oppenheimer,\cite{dboc1,dboc2,dboc3,dboc4} scalar relativistic,\cite{rel1,rel2} full triple and perturbative quadruple excitations \cite{bomble05,kallay05,kallay08}) and systematic use of nZ basis sets leads to the CBS-CVH scheme.

The effect of spin-orbit coupling is added to the energies of the O, OH, SH and Cl radicals, lowering their electronic energies by 0.22, 0.20, 0.54 and \SI{0.84}{\kilo\cal\per\mol}, respectively.\cite{moore} 

Vibrational contributions are always obtained by the rev-DSDPBEP86-D3(BJ) double-hybrid functional, \cite{santra2019minimally} 
in conjunction with the j3Z basis set (hereafter rev-DSD). Harmonic frequencies are computed by analytical second derivatives \cite{Biczysko10} and anharmonic corrections, when needed, by the generalized second order vibrational perturbation theory (GVPT2) employing third- and semi-diagonal fourth derivatives obtained by numerical differentiation of second derivatives implemented by one of the present authors in the Gaussian software \cite{Barone-JCP2005,gvpt2,g16}. 

All the computations have been performed with the Gaussian code \cite{g16}, except CCSD(T) geometry optimizations that have been carried out with the Molpro package \cite{Molpro}, CCSDT or CCSDT(Q) energy evaluations with the MRCC program \cite{mrcc} and  DBOC together with relativistic computations with the CFOUR code.\cite{CFour}

\subsection{The jChS model chemistry}
The jChS total electronic energies are obtained by single point computations at rev-DSD geometries:
\begin{equation}
    E_{\mathrm{jChS}}=E(\mathrm{CCSD(T)/j3Z}+\Delta E^{\mathrm{CBS}}_{\mathrm{MP2}}+\Delta E_{\mathrm{CV}}
\end{equation}
where the CBS term is
\begin{equation}
  \Delta E^{\mathrm{CBS}}_{\mathrm{MP2}}=\frac{4^3E(\mathrm{MP2/j4z)}
  -3^3E(\mathrm{MP2/j3Z})}{4^3-3^3}  
\end{equation}

and the core valence correction $\Delta E_{\mathrm{CV}}$ is the MP2 energy difference between all electron (ae) and frozen core (fc) calculations employing the cc-pwCVTZ basis set \cite{WD95}. At this level, the extrapolation of Hartree-Fock (HF) and correlation contributions is performed with the same equation and basis sets since several tests have shown that this simplified recipe has a negligible impact on the overall accuracy of the results. Furthermore, scalar relativistic effects are neglected, which is not a serious approximation since the heaviest element involved in this study is Cl.

\subsection{The CBS-CVH composite scheme}
The CBS-CVH total electronic energies are obtained from single-point computations at geometries optimized by the jChS composite method described above for energies:

\begin{equation}
E_{\mathrm{tot}} = E^{\mathrm{CBS}}_{\mathrm{HF}} +\Delta E^{\mathrm{CBS}}_{\mathrm{CCSD(T)}} +\Delta E_{\mathrm{CV}} +\Delta E_{\mathrm{fT}} +\Delta E_{\mathrm{pQ}} +\Delta E_{\mathrm{rel}} +\Delta E_{\mathrm{DBOC}}.
\end{equation}

In this case, HF and correlation energies are extrapolated separately. In particular, the HF CBS limit is estimated by using Feller's exponential formula \cite{Feller92}

\begin{equation}
E_{\mathrm{HF}}(n)=E_{\mathrm{HF}}^{\mathrm{CBS}}+B\exp(-Cn)
\end{equation}

whereas the CBS limit of the correlation energy is obtained by the $n^{-3}$ formula proposed by Helgaker and coworkers \cite{helgaker97}

\begin{equation}\label{eq:helgaker}
\Delta E_{\mathrm{corr}}(n)=\Delta E_{\mathrm{corr}}^{\mathrm{CBS}}+An^{-3}
\end{equation}

The three-point extrapolation of HF energies employs 3Z,4Z and 5Z basis sets, whereas the two smaller basis sets are used in the two-point extrapolation of correlation energies. 
The core valence correction $\Delta E_{\mathrm{CV}}$ is computed as the CCSD(T) energy difference between all electron and frozen core calculations employing the cc-pCVTZ basis set \cite{WD95}.

The diagonal Born-Oppenheimer correction $\Delta E_{\mathrm{DBOC}}$ \cite{dboc1,dboc2,dboc3,dboc4} and the scalar relativistic contribution to the energy 
$\Delta E_{\mathrm{rel}}$ \cite{rel1,rel2} are computed at the HF-SCF/aug-cc-pVDZ and CCSD(T)/aug-cc-pCVDZ level, after having checked their convergence with respect to contributions calculated with triple-zeta basis sets for a few stationary points. 

Finally, the corrections due to full treatment of triple ($\Delta E_{\mathrm{fT}}$) and perturbative treatment of quadruple ($\Delta E_{\mathrm{pQ}}$) excitations are computed, within the fc approximation, as energy differences between CCSDT and CCSD(T) and between CCSDT(Q) and CCSDT calculations employing the cc-pVTZ and cc-pVDZ basis set, respectively.

\subsection{Kinetic models}
Global and channel-specific rate constants were computed solving the multi-well one-dimensional master equation using the chemically significant eigenvalues (CSEs) method within  the Rice-Ramsperger-Kassel-Marcus (RRKM) approximation.\cite{georgievskii2013reformulation}. The collisional energy transfer probability is described using the exponential down model \cite{tardy1966} with a temperature dependent ${\Delta E}_{down}$ of  $260\times(T/298)^{0.875}$ cm$^{-1}$ in an argon bath gas. 

For channels ruled by a distinct saddle point, rate coefficients are determined by conventional transition state theory (TST) within the rigid-rotor harmonic-oscillator (RRHO) approximation\cite{TS_Truhlar} and including tunneling as well as non classical reflection effects by using the Eckart model. \cite{eckart1930penetration} Instead, rate constants for barrierless elementary reactions are computed employing phase space theory (PST) \cite{hunter1993,Skouteris2018}, again within the RRHO approximation. The isotropic attractive potential $V_{\mathrm{eff}}$ entering the PST is described by a $\frac{C}{R^{6}}$ power law, whose C coefficient is obtained by fitting rev-DSD energies computed at various long-range distances of fragments. We obtained the following C coefficients for the PST calculations of barrierless channels: \SI{230}{\bohr\tothe{6}\hartree} for the \ce{H2S + Cl} entrance channel, \SI{64.2}{\bohr\tothe{6}\hartree} for the \ce{CH3NH2 + CN} entrance channel on the methyl side and \SI{94.4}{\bohr\tothe{6}\hartree} for the \ce{CH3NH2 + CN} entrance channel on the nitrogen side.

The rate constants of the overall reactions evaluated in different temperature ranges are fitted by the three-parameter modified Arrhenius equation proposed by Kooij \citep{kooij,laidler96}:
\begin{equation}
\label{eq:kooij}
    k(T)=A\left(\frac{T}{300}\right)^n\exp\left(-\frac{E}{RT}\right)
\end{equation}
where $A$, $n$, and $E$ are the fitting parameters, and $R$ is the universal gas constant. 

\section{Results and discussion}
In the original jChS model, geometries and force fields were computed with the B2PLYP double hybrid functional \cite{Grimme2006} augmented by empirical dispersion contributions (namely the D3(BJ) model) \cite{grimme2011,caldeweyher2017} in conjunction with partially augmented triple-zeta basis sets.\cite{alessandrini2020}
However, the recently developed rev-DSD model \cite{santra2019minimally} delivers improved descriptions of non-covalent interactions and activation energies.\cite{xie20,alonso21} Therefore, we benchmarked the performances of this functional (still in conjunction with partially augmented triple-zeta basis sets) for geometrical parameters and vibrational frequencies obtaining results close to those delivered by the CCSD(T) ansatz in conjunction with comparable basis sets, but at much reduced computational cost \cite{Barone20}.
As a consequence, the jChS model chemistry now uses by default rev-DSD geometries and force-fields.

If the spin contamination from higher spin states is large, the potential energy surfaces computed by unrestricted wave functions can be significantly distorted showing, for example, anomalously high reaction barriers.\cite{sosa86} This means that 
UMP2 estimates of CBS and CV contributions in the jChS model could become problematic. On the other hand, CCSD fully eliminates the S+1 contaminant\cite{schlegel88} and CCSD(T) reduces also the S+2 contaminant,\cite{yuan2000} so that calculations at the CCSD(T) level are usually  relatively insensitive to the choice of (restricted or unrestricted) orbitals \cite{stanton94}. However, in cases where higher spin contaminants become important, CCSD(T) can also fail.\cite{yuan2000} On these grounds, all the jChS and CBS-CVH energies have been computed by the restricted open-shell approach. 

Concerning DFT methods, it is well known that the extent of spin contamination in unrestricted versions of hybrid density functionals increases with the amount of HF exchange \cite{cohen07}. However, Menon and Radom \cite{menon08} showed that in unrestricted double-hybrid procedures, the opposing behavior of UHF and UMP2 with respect to spin contamination leads to smaller differences between the energies predicted by unrestricted and restricted open-shell variants. Although rev-DSD energies are not used in the present context, spin contamination can have an effect also on gradients and Hessians. We have, therefore, checked systematically the spin contamination and found that its effect is always negligible (within the target accuracy of the jChS model) except for the CN radical and the transition state ruling the reaction \ce{H^{.} + F2 -> HF + F^{.}}, which will be analyzed in detail in a following section. 

\subsection{Reaction barriers}
 The most well-known database of accurate reaction barriers is the DBH24 compilation \cite{karton2008,zheng2009} 
 containing results mostly obtained at the CCSDTQ5/CBS level via W4 theory \cite{w4} for a statistically representative set including 3 prototypes for each of the following classes of reactions: 
 heavy atom transfer, nucleophilic substitution, unimolecular and association reactions, and hydrogen-transfer reactions.
 
 Table \ref{tab:barrierDBH24} compares the reaction barriers computed at CCSD(T) and jChS levels to the reference
values of ref. \citenum{zheng2009}.  The arithmetic (Mean Unsigned Error – MUE) and geometric (Root Mean Square Deviation – RMSD) average errors show that the jChS model chemistry fulfills the target of sub-chemical accuracy without any outlier above \SI{1}{\kilo\cal\per\mol} (max error = \SI{0.80}{\kilo\cal\per\mol}). It is also remarkable that estimation of CBS and CV contributions by inexpensive MP2 computations and without any empirical parameter, halves the error of the underlying CCSD(T) computation. In order to investigate the role of geometries on the computed barriers, we have repeated the computations using the QCISD/MG3 structures employed in the original compilation.\cite{zheng2009} It is quite apparent that in this case the results are only marginally affected by geometry optimizations at different computational levels. We will come back to this aspect in the following since the situation could be different for more complex transition structures and/or the non-covalent complexes ruling the entrance channels of barrier-less reactions. In this connection, further support to the reliability of rev-DSD structures is provided by the respectable MUE and RMSD (1.7 and \SI{2.4}{\kilo\cal\per\mol}, respectively) of the energy barriers computed at this level.

Zhang and co-workers \cite{dbh24_f12} have shown that, for the same set of reactions, inclusion of explicit correlation (F12) in CCSD(T) computations \cite{f12_chemrev} reduces the mean and maximum unsigned errors of the conventional CCSD(T) approach (0.66 and \SI{1.77}{\kilo\cal\per\mol}) to 0.29 and \SI{0.85}{\kilo\cal\per\mol} when using basis sets slightly larger than j3Z (including also f diffuse functions on non-hydrogen atoms). As shown in Table \ref{tab:barrierDBH24}, this improvement is close to that obtained when going from CCSD(T)/j3Z  (0.71 and \SI{2.49}{\kilo\cal\per\mol}) to jChS (0.36 and \SI{0.80}{\kilo\cal\per\mol}). These trends suggest that either inclusion of explicit correlation or two-point extrapolation at the MP2 level are effective routes for improving significantly the accuracy of computed energy barriers, without introducing additional computational bottlenecks with respect to the underlying CCSD(T)/j3Z reference. As a matter of fact, already for reactions involving two heavy atoms (e.g., A7, A8, A9, A10 in Table \ref{tab:barrierDBH24} single-point jChS computations require no more than twice the time of the CCSD(T)/jun-cc-pVTZ step and are an order of magnitude faster than the CBS-CV counterparts. Increasing the dimensions of the systems the effectiveness of the jChS model increases because of  the favorable scaling of MP2 computations with respect to CCSD(T) ones, which can be further enhanced by approaches employing resolution of identity and other acceleration techniques. Furthermore, jChS computations can be performed also with the widely diffused electronic structure codes lacking explicitly correlated approaches (e.g. Gaussian or CFOUR) and the accuracy of the results surpasses that of all the model chemistries considered by Zhang et al.\cite{zheng2009} 

\begin{table}[ht]
\centering
\caption{Theoretical values of the barrier heights in the DBH24/08 dataset obtained at different levels of theory. All the values (exclusive of ZPE) are in \SI{}{\kilo\cal\per\mol}.}
\label{tab:barrierDBH24}
\begin{tabular}{@{}cccccc@{}}
\toprule
   & reactions                & \multicolumn{4}{c}{forward/reverse barrier height}                                      \\ \midrule
   &                          & CCSD(T) & jChS & jChS\textsuperscript{a} & ref.\textsuperscript{b}\\ 
   \multicolumn{5}{c}{Heavy-Atom Transfer}  \\
a1$^c$  & \ce{H^{.} + N2O -> OH^{.} + N2}                   & 17.89/84.96     & 17.53/83.25 & 17.58/83.27 &17.13/82.47        \\
a2  & \ce{H^{.} + ClH -> HCl + H^{.}}                   & 18.89/18.89     & 17.31/17.31 & 17.33/17.33 &18.00/18.00        \\
a3$^c$  & \ce{CH3^{.} + FCl -> CH3F + Cl^{.}}               & 7.21/62.20      & 7.16/60.37  & 7.05/60.28 &6.75/60.00         \\
\multicolumn{5}{c}{Nucleophilic Substitution}\\
a4 & \ce{Cl^{-}\bond{...}CH3Cl -> ClCH3\bond{...}Cl^{-}}& 13.56/13.56     & 13.26/13.26 & 13.28/13.28 &13.41/13.41         \\
a5 & \ce{F^{-}\bond{...}CH3Cl -> FCH3\bond{...}Cl^{-}}  & 3.52/29.47      & 3.39/29.09  & 3.41/29.09 &3.44/29.42           \\
a6  & \ce{OH^{-} + CH3F -> HOCH3 + F^{-}}               & -2.39/17.78     & -2.48/17.36 & -2.51/17.35 &-2.44/17.66         \\
\multicolumn{5}{c}{Unimolecular and Association}                                                                           \\
a7  & \ce{H^{.} + N2 -> HN2^{.}}                        & 15.23/11.01     & 14.34/11.12 &14.36/11.09  &14.36/10.61         \\
a8 & \ce{H^{.} + C2H4  -> C2H5^{.}}                     & 2.43/42.59      & 1.9/42.19   & 1.92/42.21 &1.72/41.75           \\
a9 & \ce{HCN <-> HNC}                 	                & 47.45/32.77     & 47.98/33.24 &48.02/33.28  &48.07/32.82         \\
\multicolumn{5}{c}{Hydrogen Transfer}                                                                                      \\
a10$^d$  & \ce{OH^{.} + CH4 -> CH3^{.} + H2O}           & 7.05/19.05      & 6.63/20.04  & 6.52/19.94 &6.71/19.60           \\
a11$^{c,d}$  & \ce{H^{.} + OH^{.} -> H2 + ^{3}O}        & 10.38/14.62     & 11.51/13.77 & 11.42/13.78 &10.71/13.12         \\
a12$^c$  & \ce{H^{.} + H2S -> H2 + HS^{.}}              & 4.23/19.23      & 3.7/17.94   & 3.69/17.96 &3.62/17.33           \\ \midrule
    & MAX                     		                    & 2.49           & 0.80       & 0.80           &                   \\
    & MUE                                               & 0.71           & 0.36       & 0.35           &                   \\
    & RMSD                     		                    & 0.97           & 0.44       & 0.43           &                   \\ \bottomrule
\end{tabular}%

\textsuperscript{a} at QCISD/MG3 geometries; \textsuperscript{b} ref. \citenum{zheng2009}; \textsuperscript{c} spin-orbit contributions on
the reverse reaction barrier; \textsuperscript{d} spin-orbit contributions on the forward reaction barrier.
\end{table}

Two larger databases of prototypical reactions are also available for barriers related to transfers of hydrogen and non-hydrogen atoms (HTBH38 \cite{zhao2005a} and NHTBH38 \cite{zhao2005b}, respectively). However, the reaction barriers not already included in the DBH24 set have been obtained at lower computational level (W1). We have thus decided to compute at the jChS level all the reactions of the above two sets not contained in the original DBH24 compilation using both rev-DSD and the original QCISD/MG3 geometries. Whenever significant discrepancies were found, the reactions were recomputed also at the CBS-CVH level.

\begin{figure*}
        \begin{subfigure}[b]{0.49\textwidth}   
            \centering 
            \includegraphics[width=\textwidth]{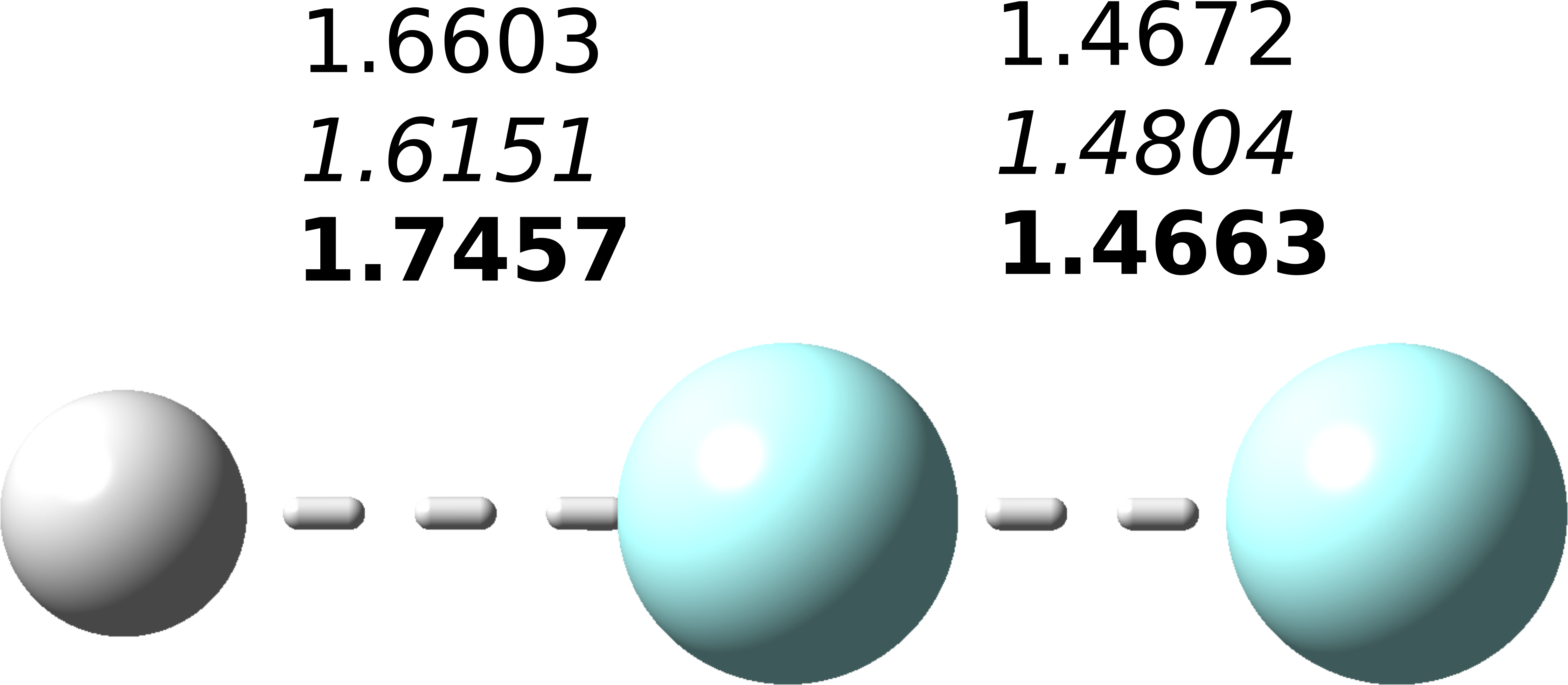}
            \caption[]%
            {{\small TS NHT3}}    
            \label{fig:tsnht3}
        \end{subfigure}
        \hfill
        \begin{subfigure}[b]{0.45\textwidth}   
            \centering 
            \includegraphics[width=\textwidth]{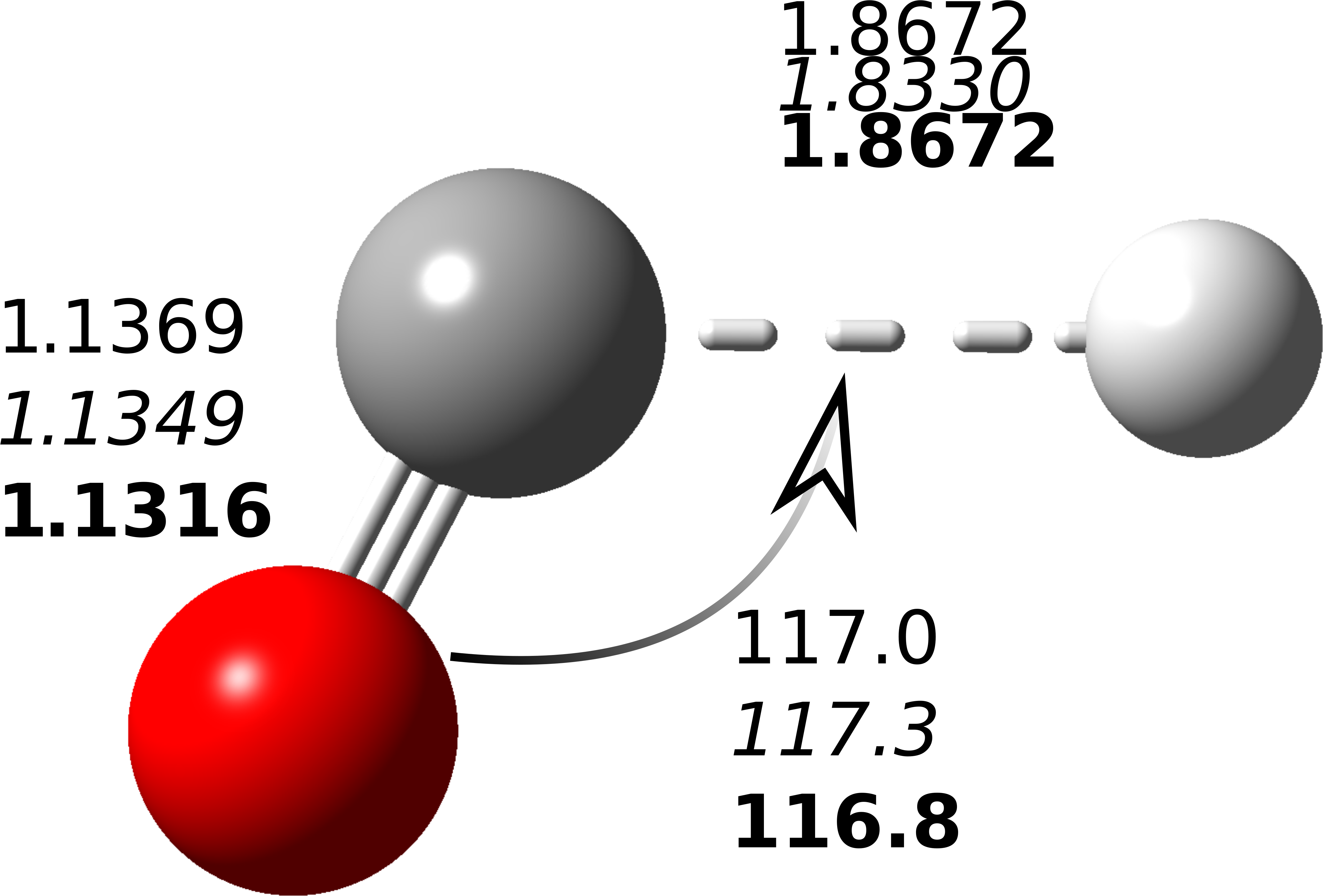}
            \caption[]%
            {{\small TS NHT9}}    
            \label{fig:tsnht9}
        \end{subfigure}
        \caption[ Dangity ]
        {\small Sketch of the structures of the transition states ruling the reactions \ce{H^{.} + F2 -> HF + F^{.}} (NHT3) and \ce{H^{.} + CO -> HCO^{.}} (NHT9). The key geometrical parameters issuing from rev-DSD, QCISD/MG3 (italics) and jChS (bold) geometry optimizations are also reported. Bond distances in \SI{}{\angstrom} and angles in degrees. The following colors are used for the different atom types: white (H),  black (C), red (O) and light blue (F).} 
        \label{fig:tsnhtbh}
    \end{figure*}

The reactions from the NHTBH38 set not included in the DBH24 selection are collected in Table \ref{tab:barrierNHTBH}. It is noteworthy that rev-DSD energy barriers, although not directly used in the jChS model chemistry, show MUEs smaller than \SI{2.0}{\kilo\cal\per\mol}, thus suggesting that the corresponding geometries should be sufficiently accurate for single-point energy evaluations at higher computational levels. This is confirmed by the finding that only for reaction NHT3, QCISD and rev-DSD geometries lead to significantly different results (cfr. columns 2 and 3 of Table \ref{tab:barrierNHTBH}). Geometry optimization at the jChS level provides results far from both values (Figure \ref{fig:tsnht3}). However, as mentioned in a previous section, unrestricted rev-DSD computations show a strong spin contamination for the TS ruling reaction NHT3 ($<$S$^2>$=1.03 in place of the correct value of 0.75). We have, therefore, re-optimized the geometry of this TS employing the restricted open-shell approach in conjunction with numerical energy derivatives. The issuing geometrical parameters (rHF=1.6603, rFF=\SI{1.4672}{\angstrom}) are closer to the jChS counterparts (rHF=1.7457, rFF=\SI{1.4663} {\angstrom}) than the unrestricted values (rHF=1.5700, rFF=\SI{1.4021}{\angstrom}) and, indeed, even better than the QCISD/MG3 values of ref. \citenum{zheng2009} (rHF=1.6151, rFF=\SI{1.4804}{\angstrom}), thus giving further support to the accuracy of rev-DSD geometries. In order to check the accuracy of computed energies irrespective of geometry effects, we have recomputed the forward and reverse barriers of reaction NHT3 at the CBS-CVH level employing QCISD/MG3 geometries. The results (1.57 and \SI{104.84}{\kilo\cal\per\mol}) are much closer to the jChS values (1.49 and 105.25, MUE=\SI{0.25}{\kilo\cal\per\mol}) than to the results of ref.\citenum{zheng2009} (2.27 and 105.80, MUE=\SI{0.83}{\kilo\cal\per\mol}) thus confirming the reliability and robustness of the jChS model chemistry. However, in this case fully reliable results can be obtained only employing more accurate geometries: as a matter of fact, the forward and reverse barriers obtained from single point CBS-CVH computations at jChS geometries are 2.59 and \SI{105.77}{\kilo\cal\per\mol}. The seemingly good agreement with the results of ref.\citenum{zheng2009} is due to a fortuitous error compensation between poor geometry and limited accuracy of the electronic energy. With the exception of this reaction the agreement between jChS energies and the reference values is satisfactory suggesting that for this kind of reactions the jChS errors are in line with those discussed above for the DBH24 database. 

\begin{table}[ht]
\centering
\caption{Theoretical values of the barrier heights for the forward and reverse reactions in the NHTBH38/08 dataset not included in the DBH24 selection. All the values (exclusive of ZPE) are in \SI{}{\kilo\cal\per\mol}.}
\label{tab:barrierNHTBH}
\resizebox{\textwidth}{!}{%
\begin{tabular}{lcccc}
\hline
    & Reaction                                           &                                & forward/reverse barrier height&                         \\ \hline
    &                                                    & jChS & jChS\textsuperscript{a} & ref.\citenum{zheng2009} \\ \cline{3-5} 
NHT1 & \ce{H^{.} + FH -> HF + H^{.}}                     & 41.99/41.99                    & 42.02/42.02                   & 42.18/42.18               \\
NHT2 & \ce{H^{.} + FCH3 -> HF + CH3^{.}}                 & 30.31/57.54                    & 30.31/57.54                   & 30.38/57.02               \\
NHT3* & \ce{H^{.} + F2 -> HF + F^{.}}                    & 3.50/107.18$^b$                & 1.49/105.25                   & 2.27/105.80               \\
NHT4 & \ce{F^{-} + CH3F -> FCH3 + F^{-}}                 & -0.70/-0.70                    & -0.71/-0.71                   & -0.34/-0.34               \\
NHT5 & \ce{F^{-}\bond{...}CH3F -> FCH3\bond{...}F^{-}}   & 13.21/13.21                    & 13.20/13.20                   & 13.38/13.38               \\
NHT6 & \ce{Cl^{-} + CH3Cl -> ClCH3 + Cl^{-}}             & 2.27/2.27                      & 2.33/2.33                     & 3.10/3.10                 \\
NHT7 & \ce{F^{-} + CH3Cl -> FCH3 + Cl^{-}}               & -12.32/19.29                   & -12.31/19.31                  & -12.54/20.11              \\
NHT8 & \ce{OH^{-}\bond{...}CH3F -> HOCH3\bond{...}F^{-}} & 11.14/47.38                    & 11.14/47.38                   & 10.96/47.20               \\
NHT9 & \ce{H^{.} + CO -> HCO^{.}}                        & 3.22/22.87                     & 3.19/22.82                    & 3.17/22.68                \\
NHT10 & \ce{CH3^{.} + C2H4 -> CH3CH2CH2^{.}}             & 6.37/32.77                     & 6.35/32.74                    & 6.85/32.97                \\ \hline
     & MAX$^c$                                           & 0.83                      & 0.80                              &                            \\
     & MUE$^c$                                           & 0.33                     & 0.32                               &                            \\
     & RMSD$^c$                                          & 0.42                      & 0.40                               &                           \\ \hline
\end{tabular}%
}
\textsuperscript{a}jChS on QCISD/MG3 geometry; \textsuperscript{b} employing restricted open-shell geometry; the values using the unrestricted geometry are: 4.46/108.14; \textsuperscript{c} neglecting the problematic reaction NHT3 (marked with an asterisk; see text for discussion).
\end{table}
%\end{landscape}

\begin{figure*}
        \centering
        \begin{subfigure}[b]{0.4\textwidth}
            \centering
            \includegraphics[width=\textwidth]{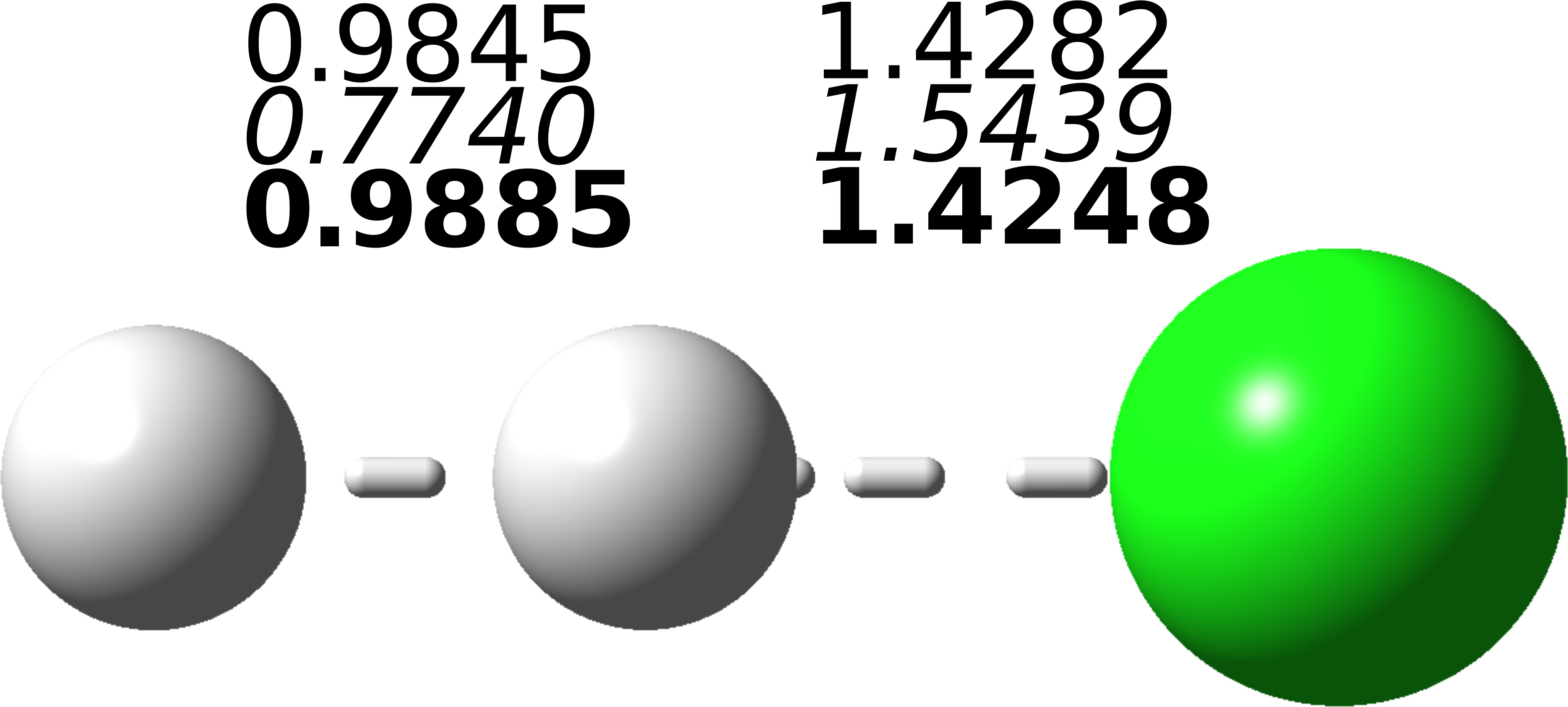}
            \caption[c1]%
            {{\small TS HT1}}    
            \label{fig:tsc1}
        \end{subfigure}
        \hfill
        \begin{subfigure}[b]{0.45\textwidth}  
            \centering 
            \includegraphics[width=\textwidth]{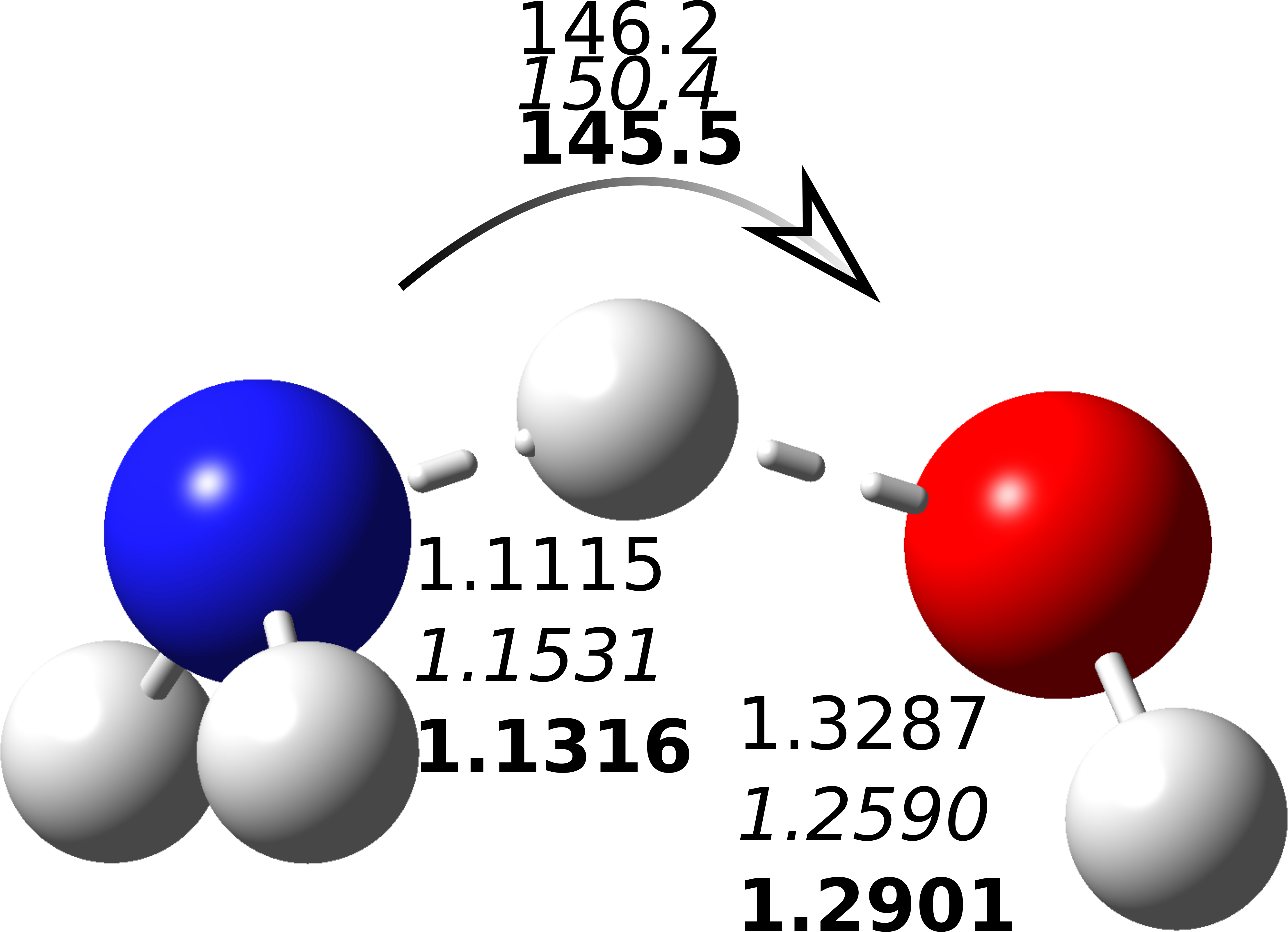}
            \caption[]%
            {{\small TS HT5}}    
            \label{fig:tsc5}
        \end{subfigure}
        \vskip\baselineskip
        \begin{subfigure}[b]{0.45\textwidth}   
            \centering 
            \includegraphics[width=\textwidth]{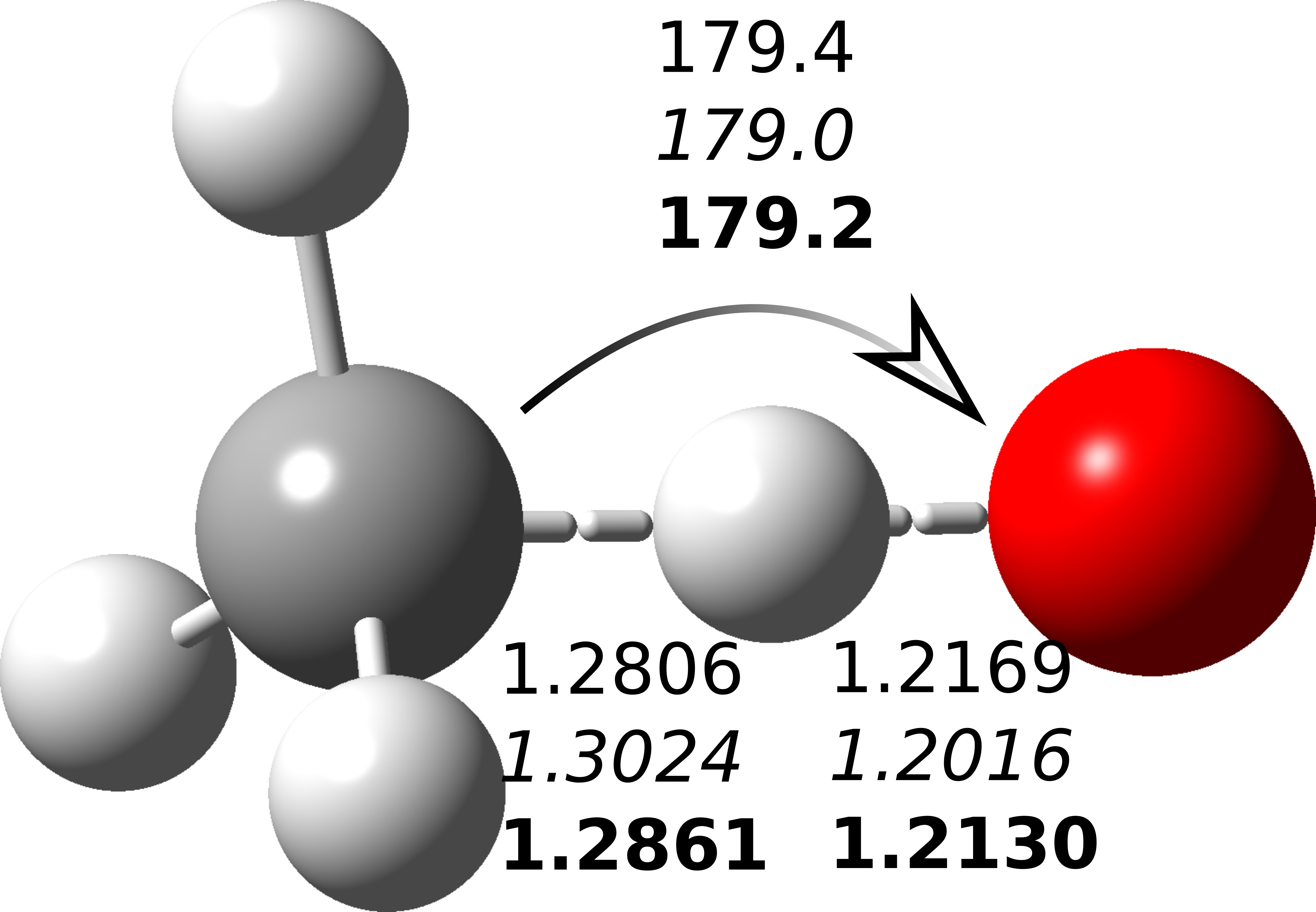}
            \caption[]%
            {{\small TS HT9}}    
            \label{fig:tsc9}
        \end{subfigure}
        \hfill
        \begin{subfigure}[b]{0.45\textwidth}   
            \centering 
            \includegraphics[width=\textwidth]{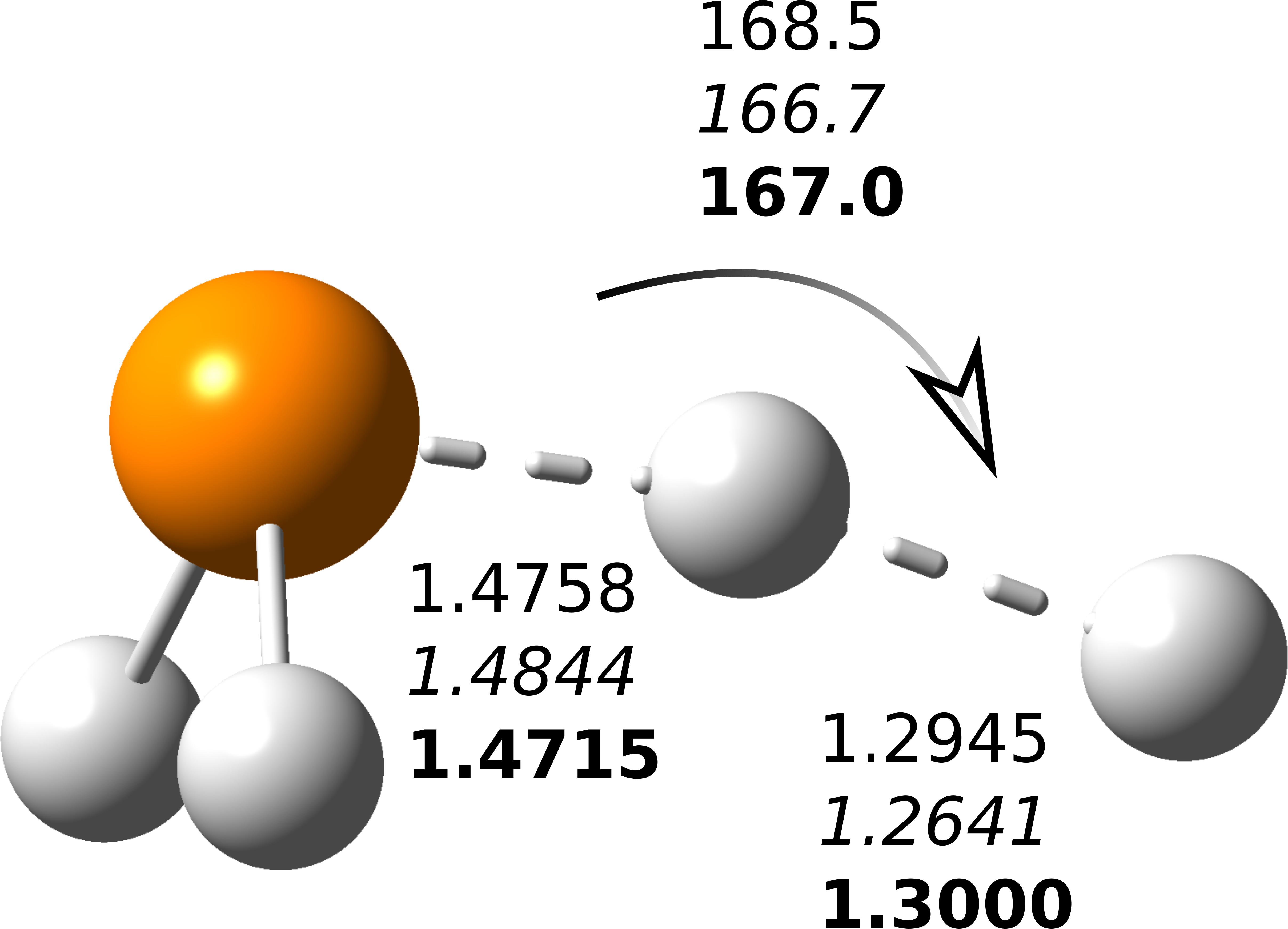}
            \caption[]%
            {{\small TS HT10}}    
            \label{fig:tsht10}
        \end{subfigure}
         \vskip\baselineskip
        \begin{subfigure}[b]{0.49\textwidth}   
            \centering 
            \includegraphics[width=\textwidth]{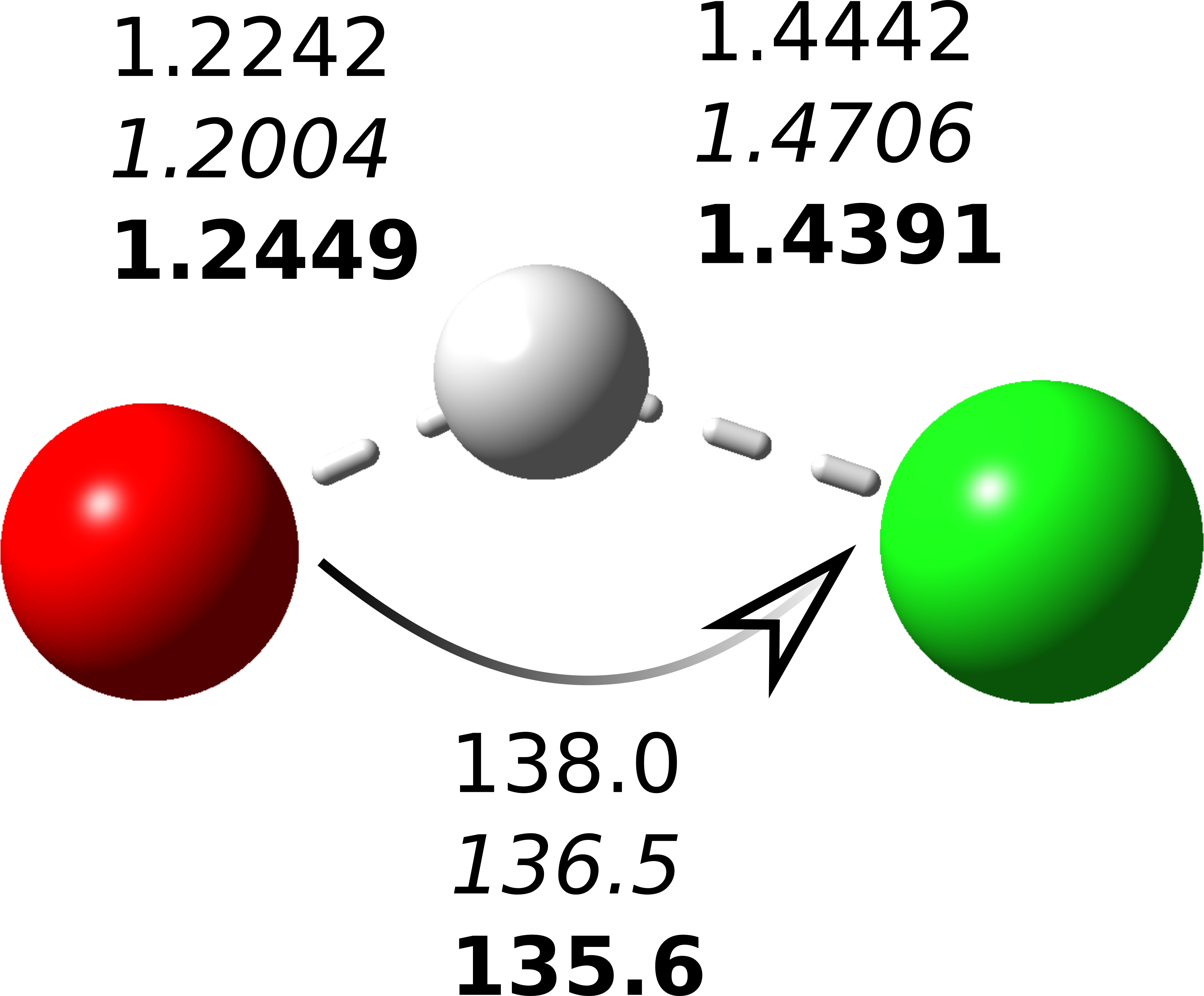}
            \caption[]%
            {{\small TS HT11}}    
            \label{fig:tsht11}
        \end{subfigure}
        \hfill
        \begin{subfigure}[b]{0.45\textwidth}   
            \centering 
            \includegraphics[width=\textwidth]{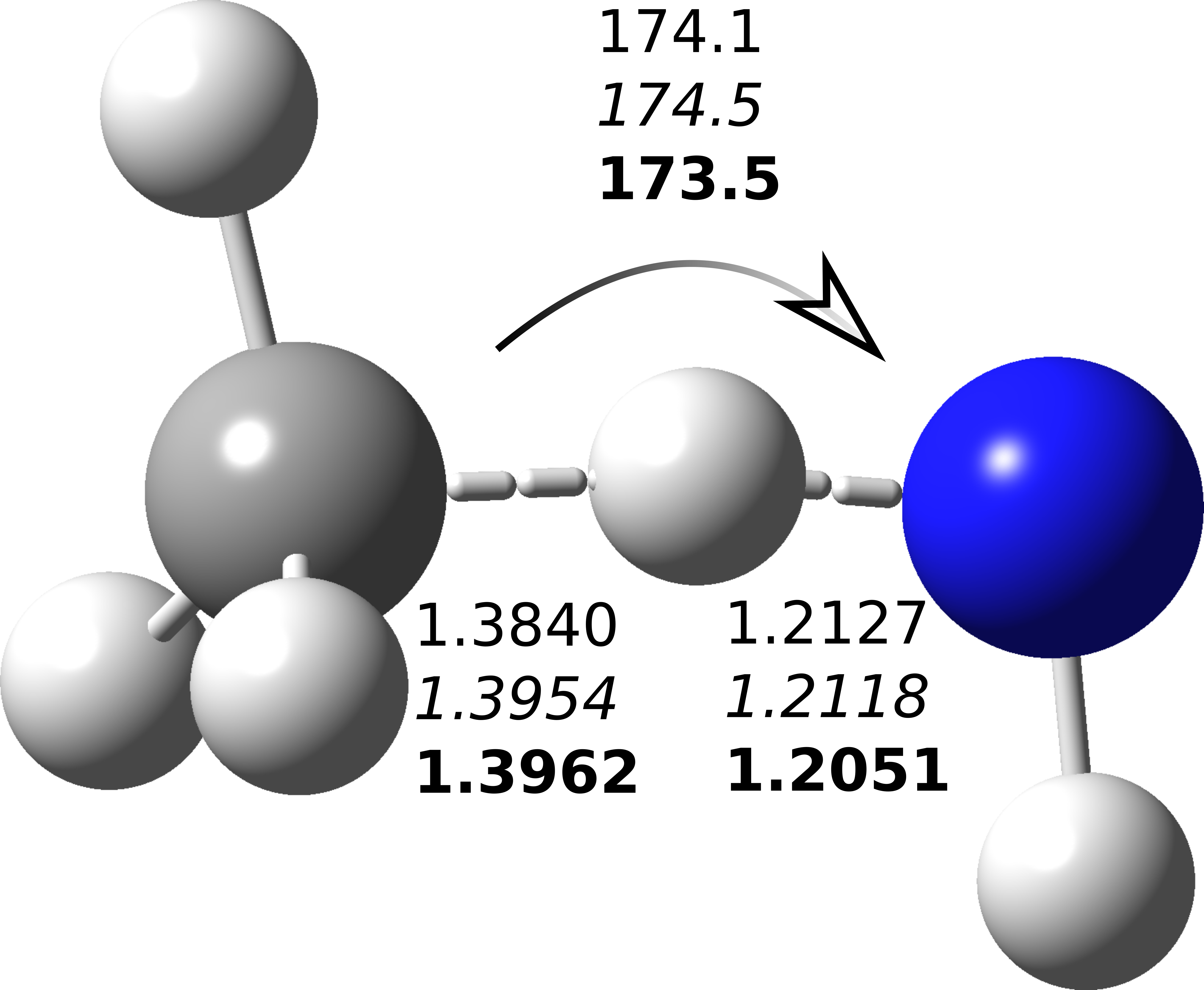}
            \caption[]%
            {{\small TS HT12}}    
            \label{fig:tsc12}
        \end{subfigure}
        \caption[ Dangity ]
        {\small Sketch of the structures of the transition states ruling the reactions collected in Table \ref{tab:paradigreac}. The key geometrical parameters issuing from rev-DSD, QCISD/MG3 (italics) and jChS (bold) geometry optimizations are also reported. Bond distances in \SI{}{\angstrom} and angles in degrees.The following colors are used for the different atom types: white (H),  black (C), blue (N), red (O), orange (P) and green (Cl).} 
        \label{fig:mean and std of nets}
    \end{figure*}

The reactions from the HTBH38 set not included in the DBH24 selection are collected in Table \ref{tab:barrierHTBH}. Once again it is noteworthy that the rev-DSD energy barriers, although not directly used in the jChS model chemistry, do not show any unrealistic outlier. Only for reactions HT1 and HT5 QCISD and rev-DSD geometries lead to significantly different results (cfr. columns 2 and 3 of Table \ref{tab:barrierHTBH}). Geometry optimization at the jChS level provides results close to rev-DSD (Figure \ref{fig:tsc1}) for HT1 and intermediate between rev-DSD and QCISD for HT5 (Figure \ref{fig:tsc5}). The agreement between jChS energies and the reference values is generally worse than for the NHTBH38 set and particularly disappointing for reactions HT1, HT5, HT9, HT10, HT11, HT12, HT13 and HT16. In order to have a first check of the accuracy of the jChS results irrespective of geometry effects, the forward and reverse barriers of two reactions in this group (HT1 and HT12) have been recomputed at the CBS-CVH level on top of QCISD/MG3 geometries. In the first case, the CBS-CVH values (5.95 and \SI{8.73}{\kilo\cal\per\mol}) are quite close to both the results of ref.\citenum{zheng2009} and the jChS counterparts for the forward barrier, but much closer to the jChS result for the reverse barrier. The situation is reversed for reaction HT12 where the CBS-CVH results (9.35 and \SI{22.37}{\kilo\cal\per\mol}) confirm the similar results of jChS and ref.\citenum{zheng2009} for the reverse barrier, but are much closer to the jChS ones for the forward barrier. Once again the jChS model chemistry does not show any outlier above the threshold of chemical accuracy, whereas this is not the case for the original reference values of ref. \citenum{zheng2009}. For the forward and reverse barriers of the remaining 8 reactions, the deviations of the jChS results from those of ref.\citenum{zheng2009} are well within sub-chemical accuracy (MUE around \SI{0.5}{\kilo\cal\per\mol}). 

\begin{table}[ht]
\centering
\caption{Theoretical values of the barrier heights for the forward and reverse reactions in the HTBH38/08 dataset not included in the DBH24 selection. All the values (exclusive of ZPE) are in \SI{}{\kilo\cal\per\mol}.}
\label{tab:barrierHTBH}
\resizebox{\textwidth}{!}{%
\begin{tabular}{lcccc}
\hline
    & Reaction                                          &      &  forward/reverse barrier height  &                                            \\ \hline
    &                                                   & jChS & jChS\textsuperscript{a} & ref. \citenum{zheng2009} \\ \cline{3-5} 
HT1*$^b$  & \ce{H^{.} + HCl -> H2 + Cl^{.}}              & 4.97/7.80                      & 5.57/7.95                     & 5.49/7.42                 \\
HT2$^c$  & \ce{OH^{.} + H2 -> H2O + H^{.}}               & 5.67/21.76                     & 5.58/21.69                    & 5.10/21.20                \\
HT3  & \ce{CH3^{.} + H2 -> CH4 + H^{.}}                  & 11.96/14.64                    & 11.95/14.64                   & 12.10/15.30               \\
HT4  & \ce{H^{.} + H2 -> H2 + H^{.}}                     & 9.58/9.58                      & 9.58/9.58                     & 9.60/9.60                 \\
HT5*$^c$  & \ce{OH^{.} + NH3 -> H2O + NH2^{.}}           & 4.13/14.41                     & 3.55/13.85                    & 3.20/12.70                \\
HT6$^b$  & \ce{HCl + CH3^{.} -> Cl^{.} + CH4}            & 1.69/7.19                      & 1.70/7.61                     & 1.70/7.90                 \\
HT7$^c$  & \ce{OH^{.} + C2H6 -> H2O + C2H5^{.}}          & 4.00/20.91                     & 3.84/20.75                    & 3.40/19.90                \\
HT8  & \ce{F^{.} + H2 -> HF + H^{.}}                     & 1.69/33.90                     & 1.77/34.00                    & 1.80/33.40                \\
HT9*$^{b,c}$  & \ce{^{3}O + CH4 -> OH^{.} + CH3^{.}}.    & 14.77/9.83                     & 14.87/9.82                    & 13.70/8.10                \\
HT10* & \ce{H^{.} + PH3 -> PH2^{.} + H2}                 & 2.85/25.09                     & 2.82/25.05                    & 3.10/23.20                \\
HT11*$^{b,c}$ & \ce{^{3}O + HCl -> OH^{.} + Cl^{.}}      & 10.81/11.38                    & 10.85/11.70                   & 9.80/10.40                \\
HT12* & \ce{NH2^{.} + CH3^{.} -> CH4 + NH}               & 9.49/22.09                     & 9.50/22.11                    & 8.00/22.40                \\
HT13* & \ce{NH2^{.} + C2H5 -> NH + C2H6}                 & 9.97/19.08                     & 10.39/19.51                   & 7.50/18.30                \\
HT14 & \ce{NH2^{.} + C2H6 -> NH3 + C2H5^{.}}             & 11.24/17.85                    & 11.18/17.80                   & 10.40/17.40               \\
HT15 & \ce{NH2^{.} + CH4 -> NH3 + CH3^{.}}               & 13.82/16.94                    & 13.80/16.92                   & 14.50/17.80               \\
HT16* & \ce{s-trans cis-C5H8 -> same}                     & 39.66/39.66                    & 39.63/39.63                   & 38.40/38.40              \\ \hline
     & MAX$^d$                                           & 1.01                        & 0.88                      &                          \\
     & MUE$^d$                                           & 0.48                        & 0.42                     &                          \\
     & RMSD$^d$                                          & 0.58                        & 0.52                         &                          \\ \hline
\end{tabular}%
}
\textsuperscript{a}jChS on QCISD/MG3 geometry; \textsuperscript{b} spin-orbit corrections on the reverse reaction barrier;
\textsuperscript{c} spin-orbit corrections on the forward reaction barrier; \textsuperscript{d} neglecting the problematic reactions (marked with an asterisk).
\end{table}

We have then selected six 'challenging reactions' among those mentioned above for further investigation. To this end, we report in Table \ref{tab:paradigreac} the results obtained at different geometries together with new reference values obtained at the CBS-CVH level on top of jChS geometries. A first general remark is that some of the new reference values differ by more than \SI{1}{\kilo\cal\per\mol} from those reported in ref.\citenum{zheng2009} (cfr. columns 1 and 6 of Table \ref{tab:paradigreac}). Furthermore, the only barrier showing significant contributions by higher order terms (mainly full triple and perturbative quadruple excitations) is the reverse barrier of reaction HT5 (cfr. columns 5 and 6 in Table \ref{tab:paradigreac}). Neglecting this barrier, the results of ref.\citenum{zheng2009} show a MUE of \SI{0.82}{\kilo\cal\per\mol} and a maximum error of \SI{1.61}{\kilo\cal\per\mol}, whereas the jChS approach has a MUE lower than \SI{0.40}{\kilo\cal\per\mol} without any absolute error larger than \SI{1}{\kilo\cal\per\mol}, irrespective of the level of the geometry optimizations. As a matter of fact, the relatively cheap rev-DSD geometries can be confidently employed for reaching sub-chemical accuracy and the use of more accurate structures does not really improve the results. The CBS-CV approach reduces significantly the MUE, but at the price of employing more accurate (and costly) geometries together with CCSD(T) computations performed with partially augmented 4Z basis sets. In conclusion, the jChS model chemistry can be confidently employed for evaluating reaction barriers of all the reactions included in the HTBH38 and NHTBH38 datasets with sub-chemical accuracy without any outlier above \SI{1}{\kilo\cal\per\mol}. 

\begin{table}[ht]
\centering
\caption{Theoretical values of the forward and reverse barriers ruling the 'challenging' HTBH38/08 reactions. All the values (exclusive of ZPE) are in \SI{}{\kilo\cal\per\mol}.}
\label{tab:paradigreac}
\resizebox{\textwidth}{!}{%
\begin{tabular}{@{}lccccccc@{}}
\toprule
\multicolumn{1}{c} {Geometry}   &   & \multicolumn{2}{c}{QCISD} & \multicolumn{1}{c}{rev-DSD} & \multicolumn{3}{c}{jChS} \\ \midrule
\multicolumn{2}{c} {forward/reverse barrier}       &ref.\citenum{zheng2009} &jChS\textsuperscript{a}  & jChS  & jChS  & CBS-CV & CBS-CVH    \\              \cmidrule(l){2-8} 
\cmidrule{2-3} HT1$^a$ & \ce{H^{.} + HCl -> H2 + Cl^{.}} & 5.49/7.42 &5.57/7.95     & 4.97/7.80     & 4.97/7.85  &5.25/8.23 & 5.41/8.19   \\
 HT5$^b$  & \ce{OH^{.} + NH3 -> H2O + NH2^{.}}           &3.20/12.70 &3.55/13.85    & 4.13/14.41    & 4.41/14.60 & 4.40/14.34& 4.39/13.63          \\ 
 HT9$^{a,b}$  & \ce{^{3}O + CH4 -> OH^{.} + CH3^{.}}         &13.70/8.10 &14.87/9.82    & 14.77/9.83    & 14.93/9.76          & 14.70/9.37& 14.64/9.30  \\ 
 HT10 & \ce{H^{.} + PH3 -> PH2^{.} + H2}                     &3.10/23.20&     2.82/25.05&2.85/25.09           & 2.85/25.12&2.87/24.48    & 2.89/24.52         \\
 HT11 & \ce{^{3}O + HCl -> OH^{.} + Cl^{.}}                  & 9.80/10.40  & 10.85/11.70         &       10.81/11.38            &    10.67/11.43       &    10.93/11.34          &     10.27/10.98              \\
 HT12 & \ce{NH2^{.} + CH3^{.} -> CH4 + NH}           &8.00/22.40 &9.50/22.11    & 9.49/22.09    & 8.94/21.84 &8.87/21.89 & 9.24/22.26  \\\midrule
% HT13 & \ce{NH2^{.} + C2H5 -> NH + C2H6}             & 7.50/18.30&10.39/19.51   & 9.97/19.08   & 9.41/18.80         & - & -   \\ \midrule              
    & MAX$^c$                                          & 1.32 &    0.84          & 0.57   &  0.60  & 0.66 &             \\
     & MUE$^c$                                         & 0.74 &    0.39          & 0.34   &  0.34  & 0.20 &             \\
     & RMSD$^c$                                        & 0.87 &    0.46          & 0.38   &  0.38  & 0.28 &             \\ \bottomrule
\end{tabular}%
}
\textsuperscript{a} spin-orbit corrections on the reverse reaction barrier; \textsuperscript{b} spin-orbit corrections on the forward reaction barrier; \textsuperscript{c} neglecting the reverse barrier of reaction HT5. 
\end{table}

\begin{figure*}
        \centering
        \begin{subfigure}[b]{0.2\textwidth}
            \centering
            \includegraphics[width=\textwidth]{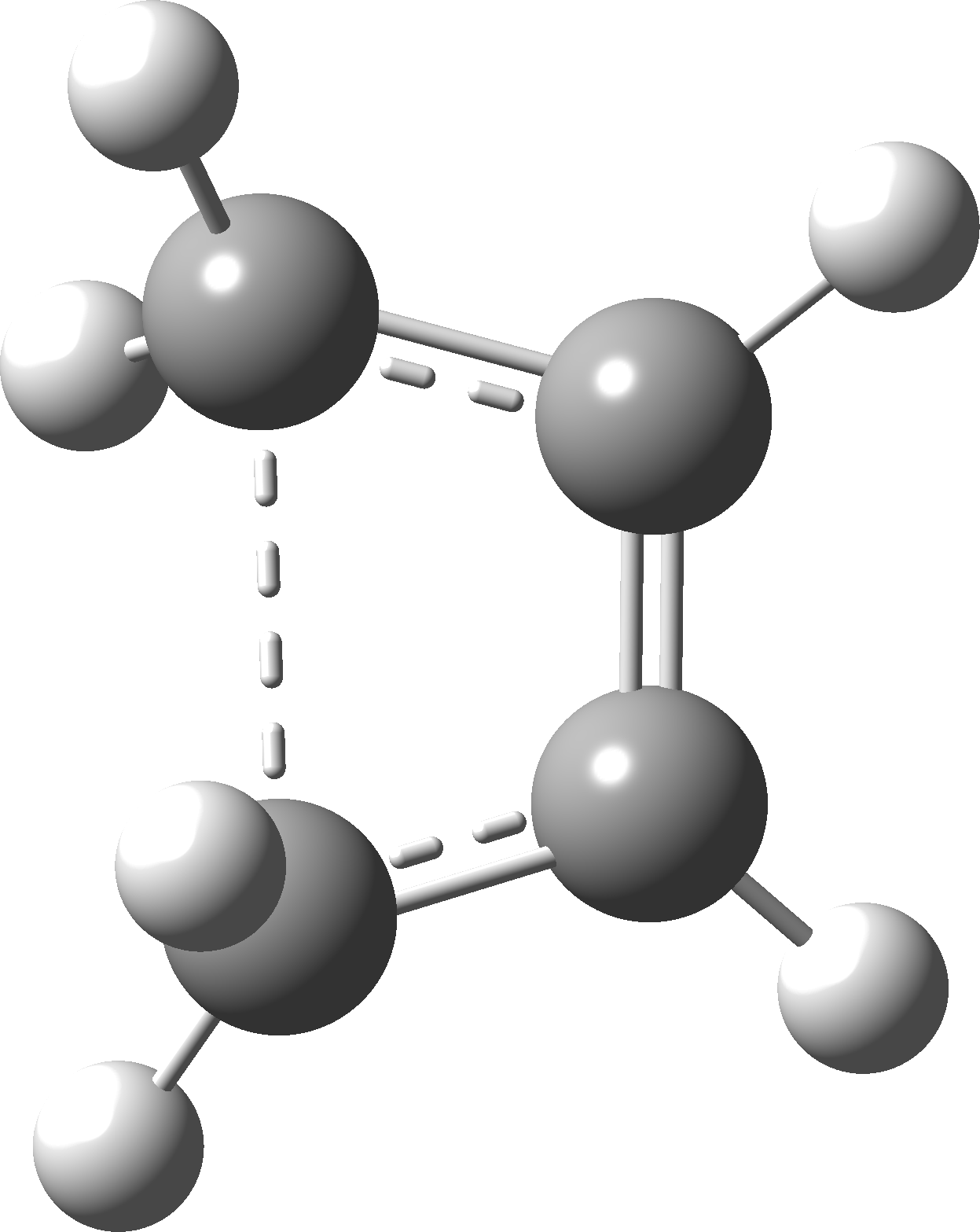}
            \caption[c1]%
            {{\small b1 BHPERI1}}    
            \label{fig:BHPERI1}
        \end{subfigure}
        \hspace{1cm}
        \begin{subfigure}[b]{0.2\textwidth}  
            \centering 
            \includegraphics[width=\textwidth]{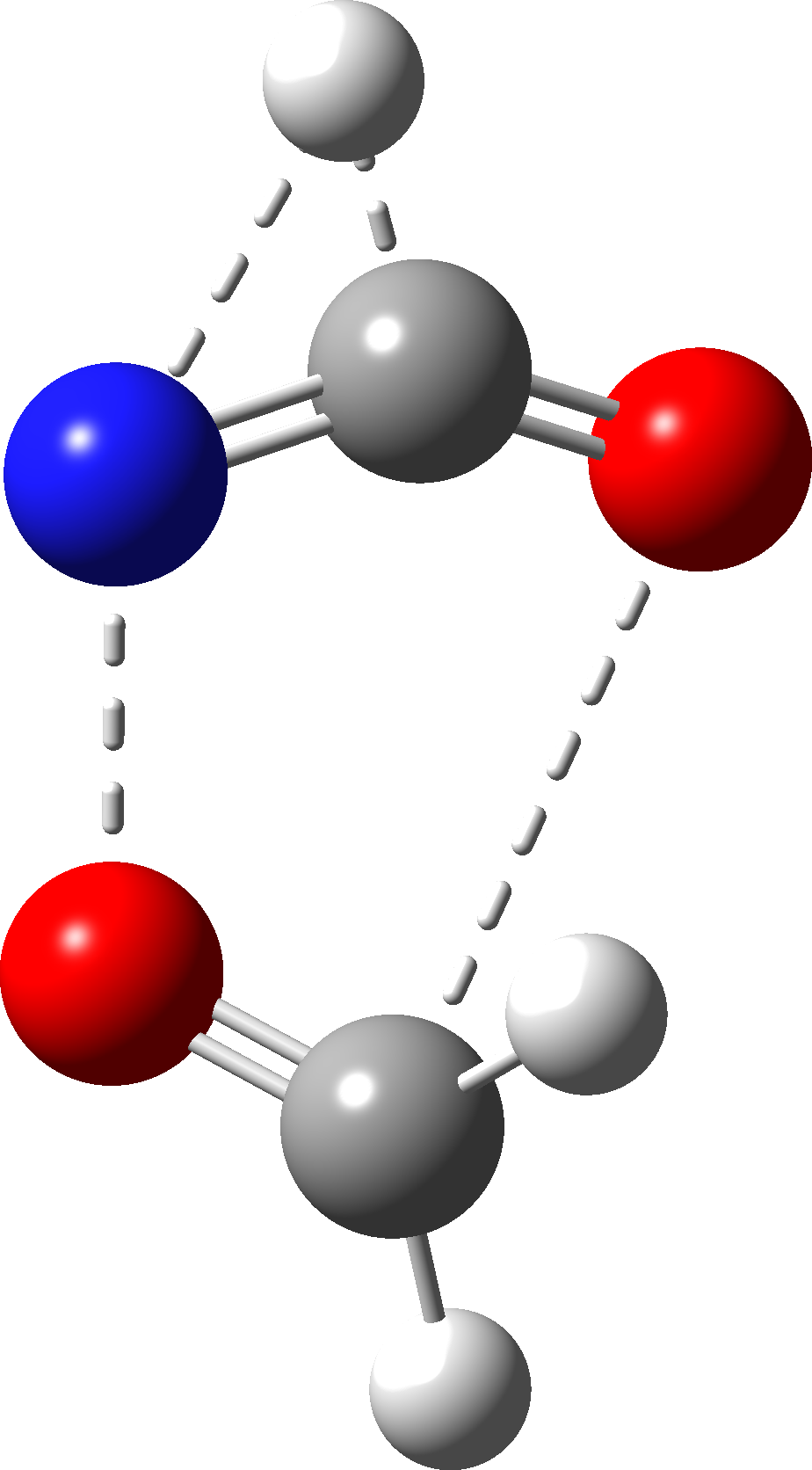}
            \caption[]%
            {{\small b2 CRBH1}}    
            \label{fig:CRBH1}
        \end{subfigure}
       \hspace{1cm}
        \begin{subfigure}[b]{0.2\textwidth}   
            \centering 
            \includegraphics[width=\textwidth]{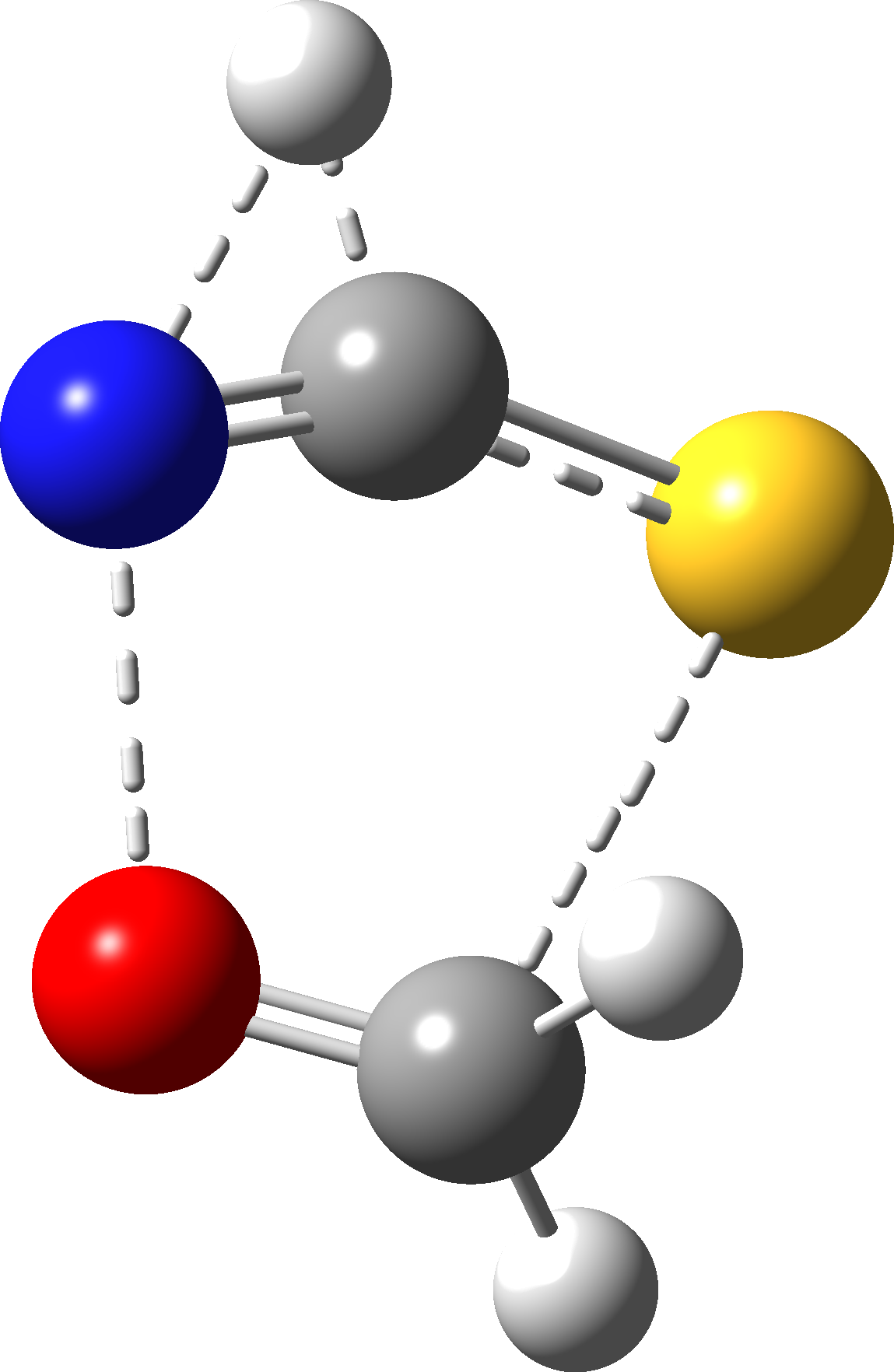}
            \caption[]%
            {{\small b3 CRBH4}}    
            \label{fig:CRBH4}
        \end{subfigure}
        \hfill
        \begin{subfigure}[b]{0.25\textwidth}   
            \centering 
            \includegraphics[width=\textwidth]{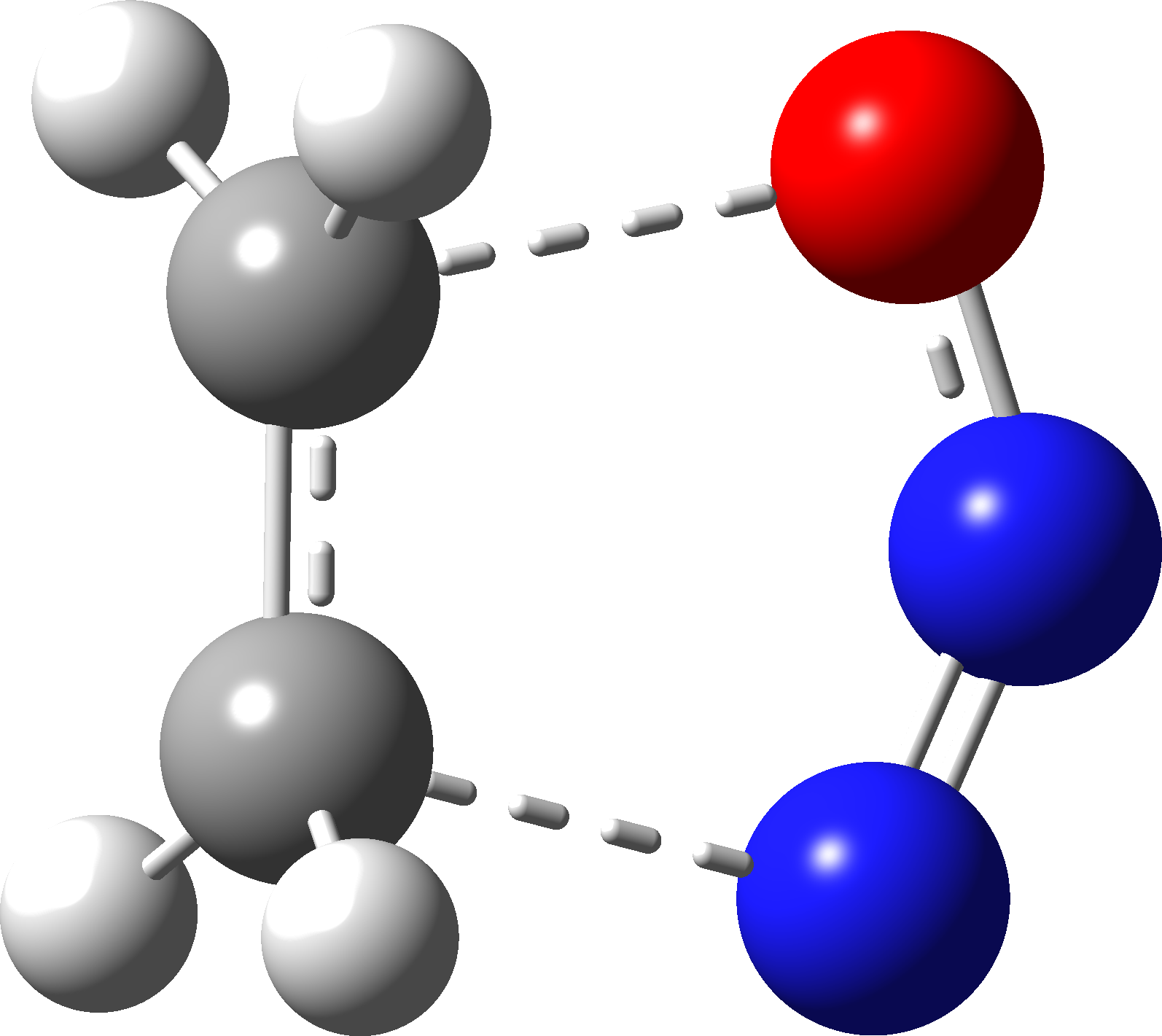}
            \caption[]%
            {{\small b4 CADBH1}}    
            \label{fig:CADBH1}
        \end{subfigure}
        \hspace{1cm}
        \begin{subfigure}[b]{0.25\textwidth}   
            \centering 
            \includegraphics[width=\textwidth]{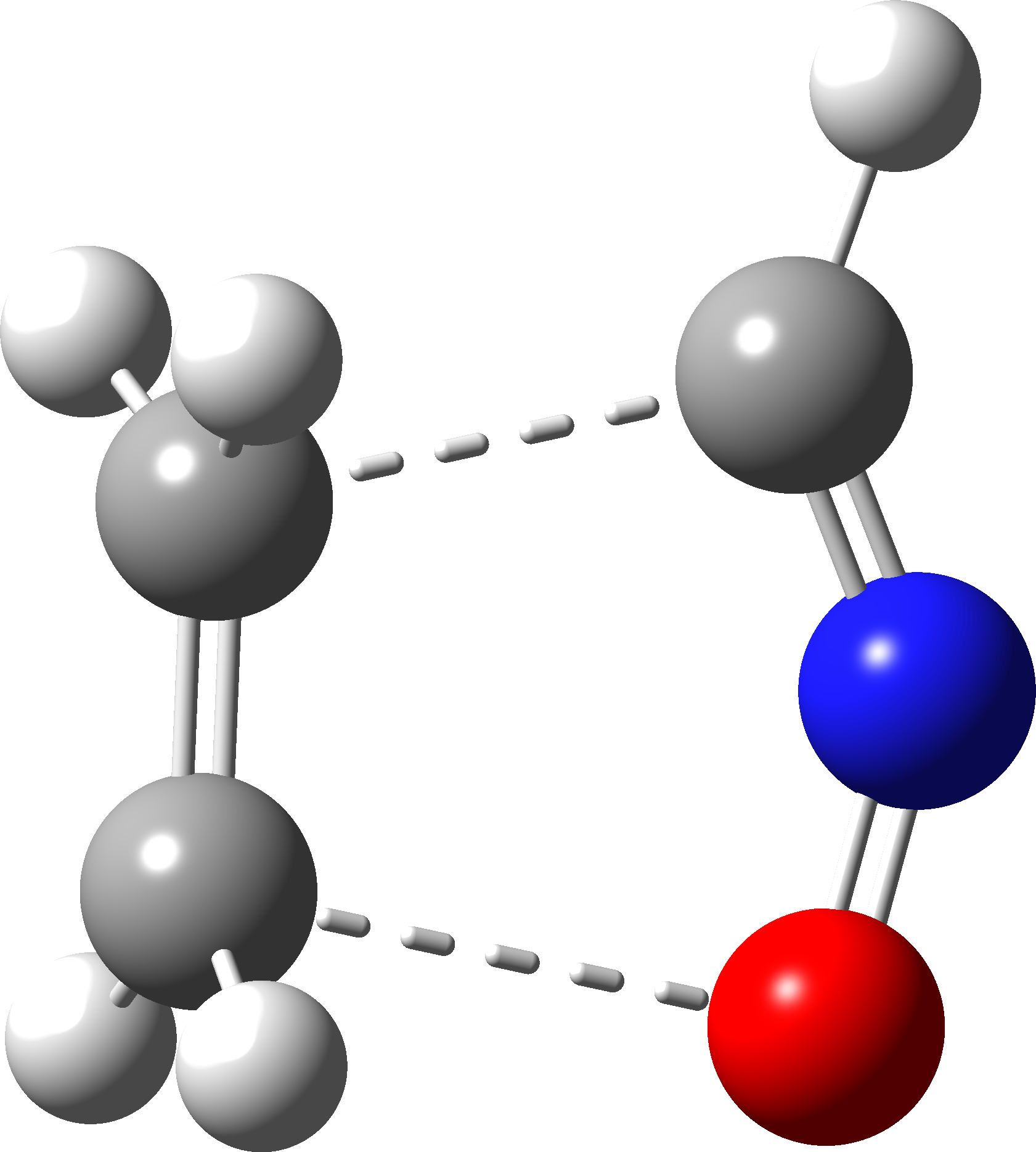}
            \caption[]%
            {{\small b5 CADBH4}}    
            \label{fig:CADBH4}
        \end{subfigure}
        \hspace{1cm}
        \begin{subfigure}[b]{0.25\textwidth}   
            \centering 
            \includegraphics[width=\textwidth]{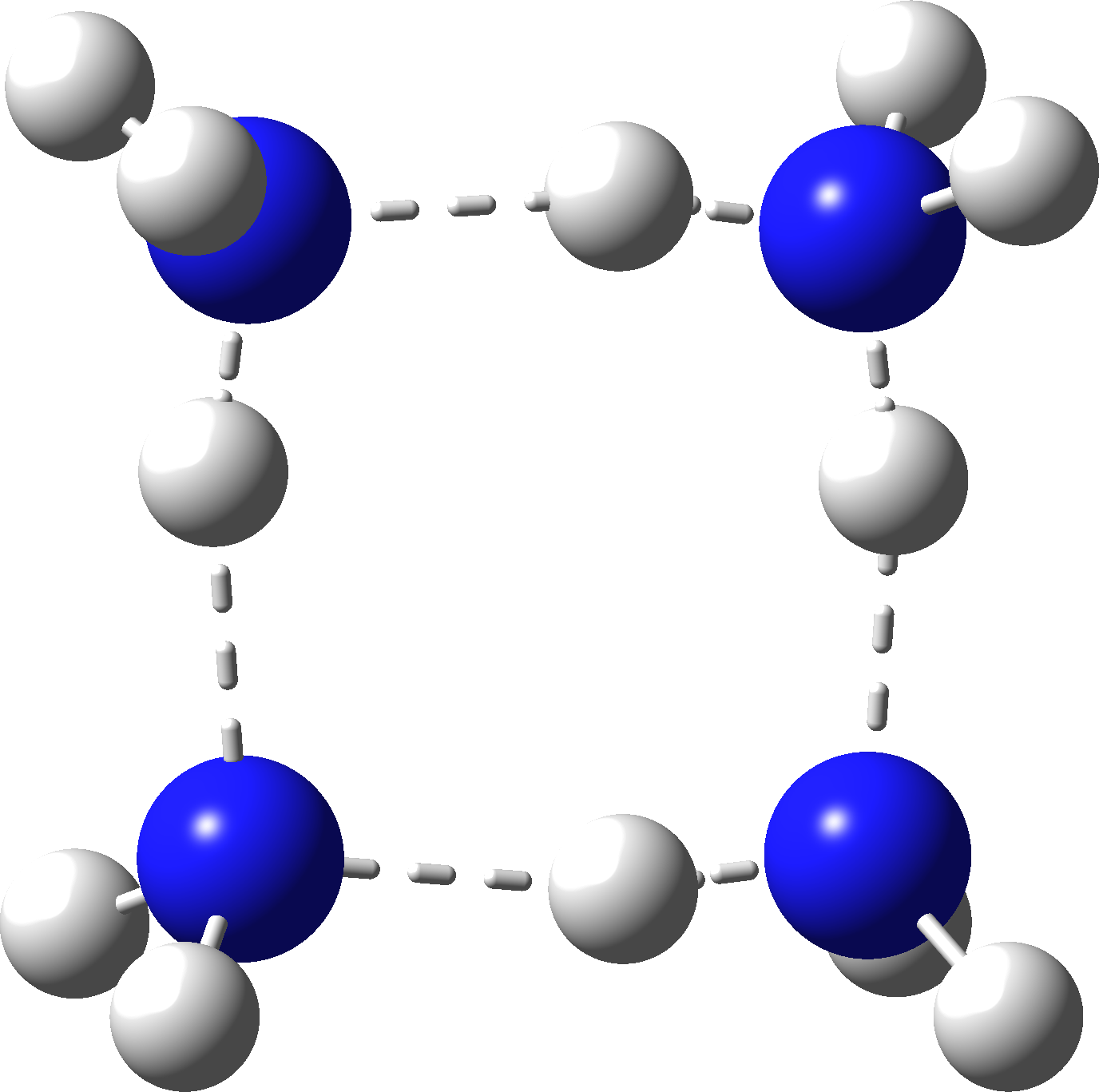}
            \caption[]%
            {{\small b6 PXBH1}}    
            \label{fig:PXBH1}
        \end{subfigure}
         \hspace{1cm}
        \begin{subfigure}[b]{0.25\textwidth}   
            \centering 
            \includegraphics[width=\textwidth]{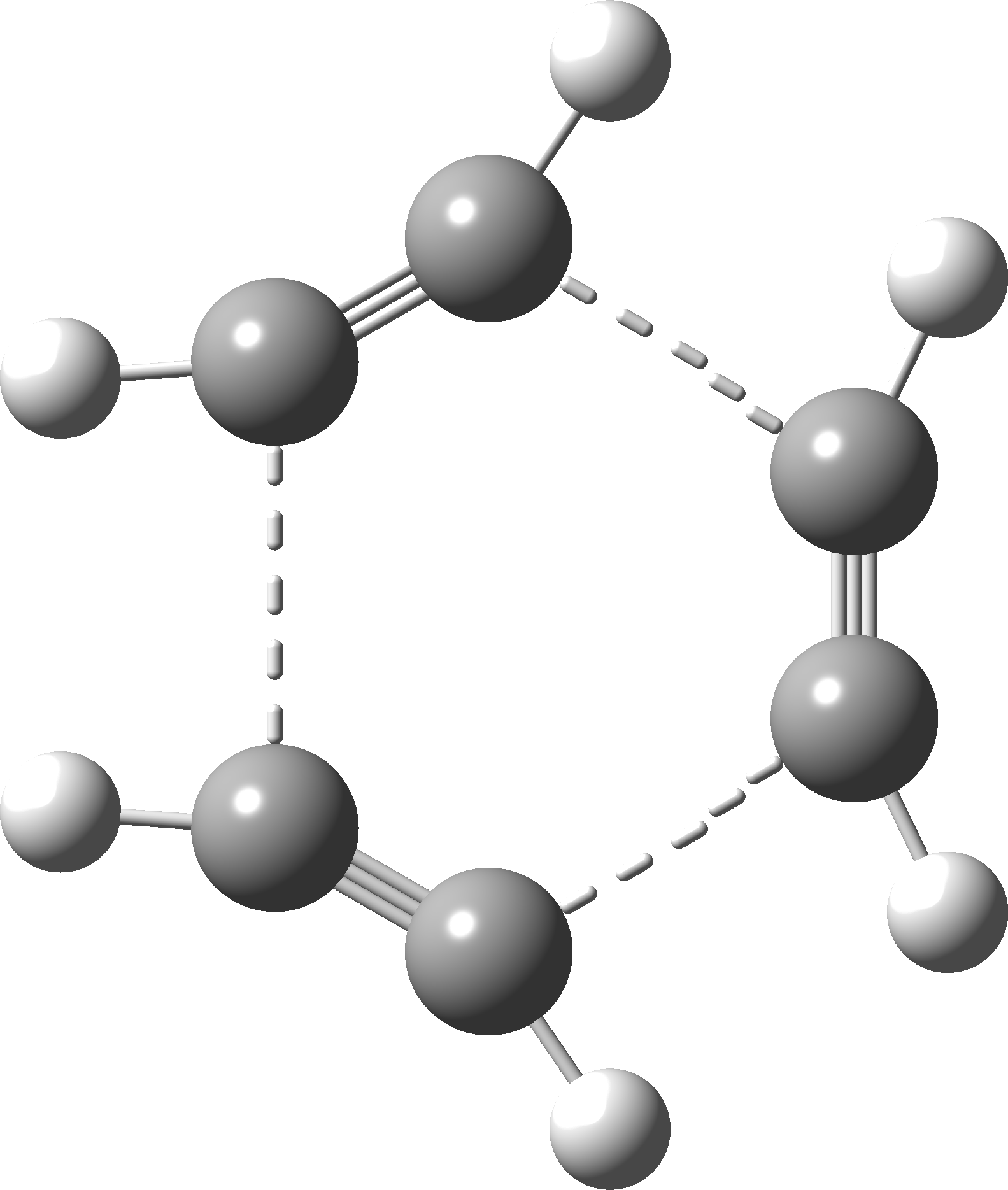}
            \caption[]%
            {{\small b7 BHDIV2}}    
            \label{fig:BHDIV2}
        \end{subfigure}
        \caption[ Dangity ]
        {Sketch of the structures of the transition states ruling the reactions of Table \ref{tab:barrierBH14}. The following colors are used for the different atom types: white (H),  black (C), blue (N), red (O) and yellow (S).} 
        \label{fig:peri}
    \end{figure*}

In order to extend the benchmark to larger and more complex systems we resorted to the BH28 set of ref.\citenum{karton2019}, which includes accurate (W3lite-F12) energy barriers for several pericyclic (BHPERI), bipolar cycloaddition (CADBH), cycloreversion (CRBH), multiple proton exchange (PXBH), and different (BHDIV) reactions. For each of those 5 classes of reactions, we selected no more than 2 representative cases. The structures of the seven selected transition states are shown in Figure \ref{fig:peri} and the corresponding forward and reverse reaction barriers (BH14 set) are collected in Table \ref{tab:barrierBH14}. 

\begin{table}[ht]
\centering
\caption{Theoretical values of the barrier heights for the forward and reverse reactions in the BH14 dataset, obtained at different levels of theory. All the values (exclusive of ZPE) are in \SI{}{\kilo\cal\per\mol}.}
\label{tab:barrierBH14}
\resizebox{\textwidth}{!}{%
\begin{tabular}{@{}cccccc@{}}
\toprule
  & label in BH28          &forward/reverse barrier height &\multicolumn{2}{c}{forward reaction barrier height}&          \\ \midrule
  &                         & jChS                      & jChS\textsuperscript{a} & W3lite-F12\textsuperscript{b} &CCSDT(Q)-CCSD(T)\\ \cmidrule(l){3-6} 
b1 & \ce{BHPERI{1}}\textsuperscript{c,d}       & 35.07/95.13                          &                 & 35.01    &-0.17   \\
b2 & \ce{CRBH{1}}\textsuperscript{e}             & 47.24 (/78.59)                          & 47.01           & 46.15  &-1.10  \\
b3 & \ce{CRBH{4}}\textsuperscript{e}            & 46.54 (/64.14)                           & 46.12           & 44.89   &-1.60     \\
b4 & \ce{CADBH{1}}\textsuperscript{d,f}       & 27.26/36.08                          &                 & 27.56    &0.00    \\
b5 & \ce{CADBH{4}}\textsuperscript{d,f} & 11.57/57.52                           &                 & 11.64    &-0.24    \\
b6 & \ce{PXBH{1}}\textsuperscript{g}            & 48.59/48.59                           &                 & 48.45   &-0.12     \\
b7 & \ce{BHDIV{2}}\textsuperscript{h}           & 51.15/201.05                           &                 & 50.10    &-0.14     \\ \midrule
   & MAX                                        &         1.65                                &                &          &         \\
   & MUE                                        &         0.62                               &                &          &         \\
   & RMSD                                       &         0.86                                  &              &          &         \\\bottomrule
\end{tabular}%
}
\textsuperscript{a} using the geometries of ref. \citenum{karton2019};\textsuperscript{b}Ref. \citenum{karton2019};\textsuperscript{c,d} Refs. \citenum{guner2003,karton2015};\textsuperscript{e}ref.\citenum{YU20151};\textsuperscript{d,f}\citenum{ess2005,karton2015};
\textsuperscript{g}ref. \citenum{karton2012b};\textsuperscript{h}ref. \citenum{goerik2017}
\end{table}

The average and maximum errors are larger than those of the DBH24 set, but, closer inspection of the results shows that, as already pointed out in ref. \citenum{karton2019}, the role of full triple and quadruple excitations is non negligible for CRBH reactions. This effect cannot be captured, of course, by the jChS model and leads to errors well above 1 \SI{}{\kilo\cal\per\mol}. In all the other cases, the errors are below the target of the jChS model chemistry. As a matter of fact, excluding the contribution of triple and quadruple excitations (last column in Table \ref{tab:barrierBH14}) reduces the MUE of jChS results to \SI{0.24}{\kilo\cal\per\mol}. Furthermore, the error related to the difference between rev-DSD and reference geometries is lower than \SI{0.3}{\kilo\cal\per\mol} even in the worst cases.

\subsection{Zero point energy and finite temperature contributions}
Accurate determination of thermochemical and kinetic parameters by quantum chemical methods requires, in addition to electronic energies, also zero point and finite temperature contributions (FTC), which are usually obtained within the RRHO approximation, possibly employing empirical scaling factors.\cite{irikura09} However, it is well known that the scaling factors are intrinsically different for zero point energies (ZPEs) and vibrational frequencies, with the results for the latter quantities being often not sufficiently accurate.\cite{freqScal15} One effective strategy devoid from any empirical parameter is offered by generalized second order vibrational perturbation theory in conjunction with a separate treatment of large amplitude motions.\cite{Barone-JCP2005,skouteris16} In fact, a resonance-free expression for ZPEs of energy minima and transition states \cite{ZPE_nonres,piccardo15}, an unsupervised smoothing procedure (HDCPT2) for fundamental frequencies \cite{hdcpt2} and a fully automatic detection and treatment of torsional motions (hindered rotor, HR, approximation) \cite{hindrot} have been implemented in the Gaussian code \cite{g16} and validated.\cite{gvpt2} As a consequence a fully black-box procedure is available for taking into account all these contributions.

Next, the so-called simple perturbation theory (SPT) \cite{spt} can be applied for computing partition functions without the need of performing explicit (or stochastic) summations of individual energy levels. In fact, the SPT retains the formal expression of the harmonic partition function, but employing the anharmonic ZPE and fundamental levels ($\Delta_i$) issuing from HDCPT2 and HR computations.

\begin{equation}
\label{eq:spt}
    Q_{vib}=\frac{\exp\left(-\frac{ZPE}{KT}\right)} {\prod_i \left[ 1-\exp\left(-\frac{\Delta_i}{KT}\right) \right]}
\end{equation}

This approximation provides results in remarkable agreement with accurate reference values and leads to analytical expressions for the different thermodynamic functions.\cite{spt}

On these grounds, we will now analyze the performances of the jChS model chemistry in dealing with these terms starting from a benchmark of the RRHO approximation with reference to accurate quantum chemical results and then proceeding to take into account anharmonic contributions. For illustration purposes,  we will focus our attention on ZPEs and absolute entropies (S), which are especially sensitive to high and low-frequencies, respectively. 

To this end a new database has been built (ThCS21), which contains accurate experimental values for the ZPEs and absolute entropies of 21 semi-rigid closed-shell molecules, whose estimated errors are below 0.1 \SI{}{\kilo\cal\per\mol} and 0.05 cal/(mol K), respectively. The results collected in Table \ref{tab:thermo21} show that already at the harmonic level, the errors are well within the level of accuracy expected from the jChS model chemistry and the anharmonic results can be confidently used in conjunction with the most sophisticated models (e.g. CBS-CVH). Actually, the harmonic frequencies obtained at this level do not require any empirical correction to compensate for method and/or basis set deficiency, but only for genuine anharmonic effects, which, in turn, give significant contributions to ZPEs only for some XH bonds (X=C,N,O). As a consequence, an empirical correction of 0.12 \SI{}{\kilo\cal\per\mol} for each bond 
of this kind provides results very close to the anharmonic counterparts (see results in parenthesis in the first column of Table \ref{tab:thermo21}).

\begin{table}[h!]
\centering
\caption{ThCS21 database: ZPEs in \SI{}{\kilo\cal\per\mol} and absolute entropies at 298.15 K and 1 atm in cal /(mol K).}
\label{tab:thermo21}
\begin{tabular}{@{}cccccc@{}}
\toprule
 Molecule    & ZPE$_{harm}^a$ & ZPE$_{anh}^{a,b}$ & ZPE$_{exp}^c$ & S$_{harm}^a$ & S$_{exp}^d$ \\ \midrule
   HF        & 5.89           & 5.84              & 5.86          & 41.46        & 41.50       \\
   HCl       & 4.30           & 4.27              & 4.24          & 44.57        & 44.64       \\  
   H$_2$     & 6.36 (6.24)    & 6.30              & 6.23          & 31.13        & 31.20       \\
   N$_2$     & 3.33           & 3.32              & 3.36          & 45.77        & 45.77       \\
   F$_2$     & 1.42           & 1.41              & 1.30          & 48.33        & 48.44       \\
   CO        & 3.09           & 3.08              & 3.09          & 47.24        & 47.21       \\
   Cl$_2$    & 0.81           & 0.81              & 0.80          & 53.18        & 53.29       \\
   CO$_2$    & 7.26           & 7.23              & 7.30          & 51.09        & 51.07       \\
   CS$_2$    & 4.36           & 4.35              & 4.34          & 56.78        & 56.85       \\
   H$_2$O    & 13.45 (13.21)  & 13.24             & 13.26         & 45.09        & 45.10       \\
   H$_2$S    & 9.59           & 9.47              & 9.48          & 49.12        & 49.16       \\
   HOF       & 8.77 (8.65)    & 8.64              & 8.65          & 54.11        & 54.17       \\
   HOCl      & 8.31 (8.19)    & 8.19              & 8.19          & 56.47        & 56.49       \\
   N$_2$O    & 6.84           & 6.80              & 6.77          & 52.51        & 52.54       \\
   HCN       & 10.03 (9.91)   & 9.95              & 10.00         & 48.16        & 48.21       \\
   SO$_2$    & 4.33           & 4.31              & 4.41          & 59.35        & 59.30       \\
 C$_2$H$_2$  & 16.72 (16.48)  & 16.56             & 16.49         & 47.91        & 47.99       \\
 H$_2$CO     & 16.76 (16.52)  & 16.54             & 16.52$^e$     & 52.23        & 52.30       \\
 NH$_3$      & 21.63 (21.27)  & 21.26             & 21.20         & 45.98        & 46.04       \\
 CH$_4$      & 28.20 (27.72)  & 27.79             & 27.71         & 44.48        & 44.48       \\
 C$_2$H$_4$  & 32.06 (31.58)  & 31.67             & 31.46$^f$     & 52.35        & 52.39       \\ \midrule
   MUE       &  0.15 (0.04)   &  0.05             &       -        &     0.05         &     -        \\
  RMSD       &  0.22 (0.06)   &  0.07             &       -        &     0.06         &     -        \\ \bottomrule
\end{tabular}%

\textsuperscript{a}revDSD-PBEP86-D3(BJ)/jun-cc-pV(T+d)Z;
\textsuperscript{b}HDCPT2 model;
\textsuperscript{c}From ref.\citenum{irikura09};
\textsuperscript{d}From ref.\citenum{nist}. The original values have been lowered by 0.03 cal/(mol K) to take into account the passage from 1 bar (0.1 MPa) to 1 atm (0.10135 MPa) references. 
\textsuperscript{e}From accurate diffusion Monte Carlo computations \cite{harding17} since the value of 16.10 reported in ref. \citenum{irikura09} is affected by an estimated error of 0.51 \SI{}{\kilo\cal\per\mol};
\textsuperscript{f} From the accurate computations of ref. \citenum{avila11} since the value of 30.70 reported in 
ref. \citenum{irikura09} is affected by an estimated error of 0.40 \SI{}{\kilo\cal\per\mol}.
\end{table}

Accurate entropy values are also available for the same set of molecules and harmonic computations perform a remarkable job in reproducing the experimental values. However, entropy is exquisitely sensitive to low-frequency vibrations, so that a set of flexible molecules is collected in Table \ref{tab:flexi}. It is apparent that the HRHO model (which does not add any computational burden with respect to the underlying RRHO model) performs a remarkable job for systems containing a single torsion. The situation is more involved for larger flexible systems 
due to the presence of several low-energy minima contributing to the overall thermodynamic functions. Although this aspect goes beyond the main topic of the present contribution, we point out that several strategies are being proposed, following systematic search \cite{li15}, 
stochastic \cite{chandramouli19} and, more recently, machine learning \cite{mancini20} approaches. Other kinds of large amplitude motions can be taken into account by means of one-dimensional variational or quasi-variational approaches \cite{puzzarini2019} followed by SPT or direct count of energy levels.\cite{skouteris16} 

\begin{table}[h!]
\centering
\caption{Absolute entropies at 298.15 K and 1 atm in cal /(mol K).}
\label{tab:flexi}
\begin{tabular}{@{}cccc@{}}
\toprule
 Molecule     & S$_{harm}^a$ & S$_{HR}^{a,b}$ & S$_{exp}$ \\ \midrule
 CH$_3$CH$_3$ & 54.38  & 54.70  & 54.79$^{c,d}$ \\
 CH$_3$OH     & 57.00  & 57.36  & 57.29$^{c,d}$ \\
 CH$_3$SH     & 60.57  & 60.99  & 60.96$^{c,d}$ \\
 CH$_3$CHO    & 62.66  & 63.11  & 63.06$^{c,d}$ \\
 CHOCHO       & 64.93  & 65.09  & 65.10$^{c,e}$ \\ \bottomrule

\end{tabular}%

\textsuperscript{a}revDSD-PBEP86-D3(BJ)/jun-cc-pV(T+d)Z;
\textsuperscript{b}including HR correction;
\textsuperscript{c}the original values have been lowered by 0.03 cal/(mol K) to take into account the passage from 1 bar (0.1 MPa) to 1 atm (0.10135 MPa) references. 
\textsuperscript{d}from ref.\citenum{chao86};
\textsuperscript{e}from ref.\citenum{dorofeeva01}.
\end{table}

Another issue is represented by open-shell species, which are of paramount importance in both astrochemistry and atmospheric chemistry.
In this case, experimental zero point energies are available only for diatomic species and accurate determinations are quite limited also for the other thermodynamic functions. 
The jChS results collected in Table \ref{tab:radicals} for a few representative systems (ThOS10 database) suggest that (in the absence of strong multi-reference effects) the expected accuracy is close to that reached for closed-shell systems.

\begin{table}[h!]
\centering
\caption{ThOS10 Database: ZPEs and non potential energy terms for representative open-shell species at 298.15 K and 1 atm in cal /(mol K).}
\label{tab:radicals}
\begin{tabular}{@{}ccccccc@{}}
\toprule
 Molecule    & ZPE$_{calc}^{a,b}$  & ZPE$_{exp}^b$ & S$_{calc}^{a,c}$ & S$_{exp}^{c,d}$ & H-H$^{0,a,b}_{calc}$ & H-H$^{0,b,e}_{exp}$ \\ \midrule
 OH($^2\pi$) & 5.25 (5.33)     & 5.29$^e$    & 43.95        & 43.88       & 2.07               & 2.11              \\
 SH($^2\pi$) & 3.88 (3.88)     & 3.82        & 47.27        & 46.76       & 2.07                   & 2.07                   \\
 CN($^2\Sigma^{+,f}$) & 2.83 (2.83)     & 2.95        & 48.35        & 48.43       & 2.07               & 2.07              \\
 NO($^2\pi$) & 2.80 (2.77)     & 2.71        & 50.42        & 50.34       
 & 2.07                    & 2.07                   \\
 NH$_2$($^2$B$_1$)      & 11.89 (11.83)   &  11.52$^g$           & 46.49        & 46.54       & 2.37               & 2.37              \\
 HCO($^2$A')         & 8.06 (8.09)     &  8.09$^h$           & 53.58        & 53.66       & 2.39                  & 2.39$^h$             \\
 HO$_2$($^2$A")      & 8.85 (8.87)     & 8.78$^i$    & 54.67        & 54.76       & 2.39               & 2.39              \\
 CH$_3$($^2$A$_2$")      & 18.62 (18.42)   & 18.48$^i$   & 46.26        & 46.38       & 2.46               & 2.45              \\
 t-HOCO($^2$A')  & 13.00 (13.07)   & 13.10$^i$   & 60.08        &  /          & 2.61               &  /                \\ 
 CH$_3$CO($^2$A')    & 26.82 (26.85)   & 26.69$^i$   & 64.23        & 63.92       & 2.98               & 2.96              \\ \bottomrule

\end{tabular}%

\textsuperscript{a}revDSD-PBEP86-D3(BJ)/jun-cc-pV(T+d)Z HRHO model and (in parenthesis) HRHO model including
a correction to ZPEs of -0.12 of for each CH, NH, or OH bond;
\textsuperscript{b} in \SI{}{\kilo\cal\per\mol};
\textsuperscript{c} in cal/(mol K);
\textsuperscript{d}from ref.\citenum{ruscic05}. When needed, entropy values have been lowered by 0.03 cal/(mol K) 
to take into account the passage from 1 bar (0.1 MPa) to 1 atm (0.10135 MPa) references;
\textsuperscript{e}From ref.\citenum{irikura09};
\textsuperscript{f} Restricted open-shell with equilibrium bond length of \SI{1.179}{\angstrom}; the unrestricted result is \SI{3.43}{\kilo\cal\per\mol} with S$^2$=0.854 and equilibrium bond length of \SI{1.159}{\angstrom};
\textsuperscript{g}CBS-CV results from ref.\citenum{demaison03}.
\textsuperscript{g}CBS-CV results from ref.\citenum{marenich03}.
\textsuperscript{i}Diffusion Monte Carlo results from ref.\citenum{harding17}.
\end{table}

\subsection{Reaction rates}
In this  section we analyze the impact on reaction rates of the different ingredients discussed in the previous section, comparing the results issuing from different model chemistries including CBS-QB3, jChS and CBS-CVH. Starting from simple elementary mechanisms we proceed to more complex potential energy surfaces including several intermediates and transition states, possibly leading to different products.

The first test case is the high pressure limit of the reaction \ce{H^{.} + CO}, which has been recently investigated by Vichietti et al. \cite{hco20}. This reaction belongs to the HTBH38 set, whose jChS results have been discussed in the Section devoted to energy barriers. For purposes of comparison we have computed also the barriers at the CBS-CVH level on top of jChS geometries obtaining values (3.26 and \SI{22.86}{\kilo\cal\per\mol}) for the forward and reverse barrier very close to the jChS counterparts at rev-DSD geometries (3.22 and \SI{22.87}{\kilo\cal\per\mol}). Although the presence of a van der Waals pre-reactive complex has been suggested, its stability (if any) is so small that its impact on the computed reaction rates is negligible.

\begin{figure}[ht]
         \centering
         \includegraphics[width=0.6\textwidth]{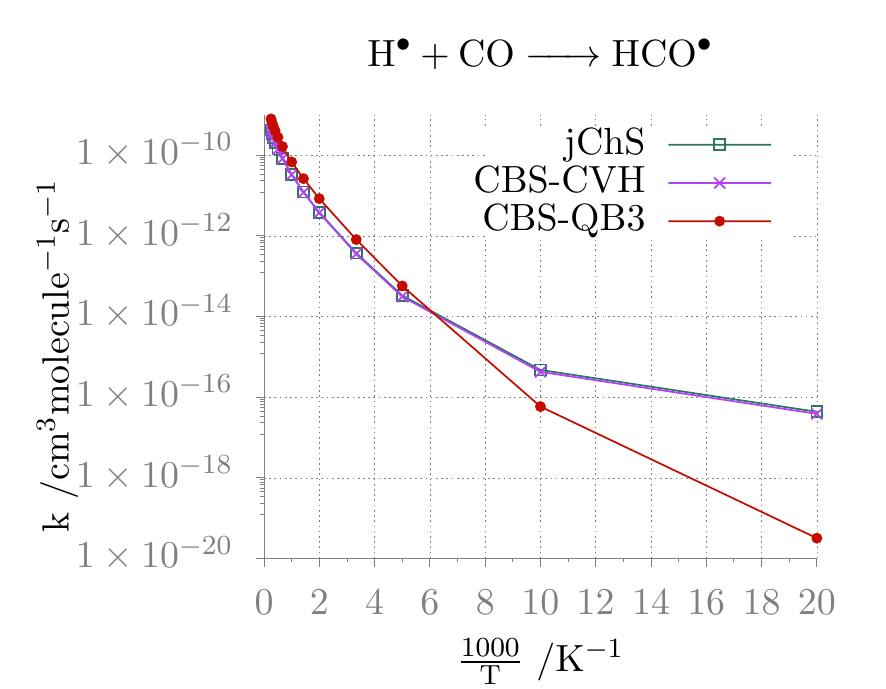}
         \caption{Temperature-dependence of the \ce{H^{.} + CO} reaction rate constants calculated at various levels of theory in the high pressure limit.}
         \label{fig:ratehco}
\end{figure}

The reaction rates computed in the 50-4000 K temperature interval are shown in Figure \ref{fig:ratehco} and the parameters of the corresponding Arrhenius-Kooij fits obtained by different electronic structure methods are collected in Table \ref{tab:arrkooijHCO}. The non-Arrhenius behaviour of the reaction is quite apparent, but the small errors of all the fits show that the Arrhenius-Kooij model captures the essential of the deviation. Furthermore, the jChS results are close to the reference values of ref.\citenum{hco20}, whereas this is not the case for the largely employed CBS-QB3 approach at least at low temperatures.

\begin{table}[ht]
\centering
\caption{The Arrhenius–Kooij parameters for the \ce{H^{.} + CO} reaction.}
\label{tab:arrkooijHCO}
\resizebox{\textwidth}{!}{%
\begin{tabular}{@{}lcccc@{}}
\toprule
 forward/reverse   & ref.\citenum{hco20} & jChS   & CBS-CVH & CBS-QB3   \\ \midrule
$A$/\SI{}{\cubic\centi\metre\per\molecule\per\second}   &\num{2.98d-11}/\num{1.37d13} & \num{3.86d-11}/\num{1.47d13}  & \num{3.87d-11}/\num{1.46d13}  & \num{7.96d-11}/\num{3.81d13}  \\
$n$   & 1.03/1.06 &\num{1.07}/\num{1.20}  & \num{1.06}/\num{1.20}  & \num{1.02}/\num{1.04}  \\
$E$/\SI{}{\kilo\cal\per\mol}   & 2.64/17.79 & \num{2.86}/18.14 & \num{2.89}/18.08 & \num{2.76}/17.95 \\
rms & - &\num{4d-14}/\num{3.20d-2}  & \num{3.9d-14}/\num{3.21d-2}  & \num{1.91d-14}/\num{2.56d-2}  \\ \bottomrule
\end{tabular}
}
\end{table}

\begin{figure*}
\begin{subfigure}[b]{0.45\textwidth}   
            \centering 
            \includegraphics[width=\textwidth]{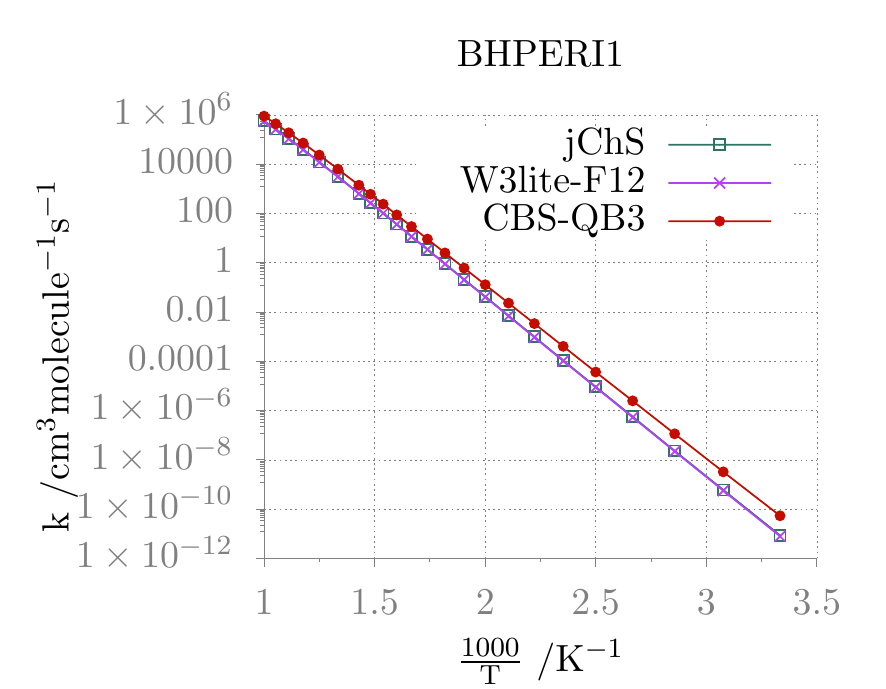}
            \caption[]%
            {{\small BHPERI1 $k(T)$}}    
            \label{fig:ratebhperi1}
        \end{subfigure}
         \hspace{1cm}
        \begin{subfigure}[b]{0.45\textwidth}   
            \centering 
            \includegraphics[width=\textwidth]{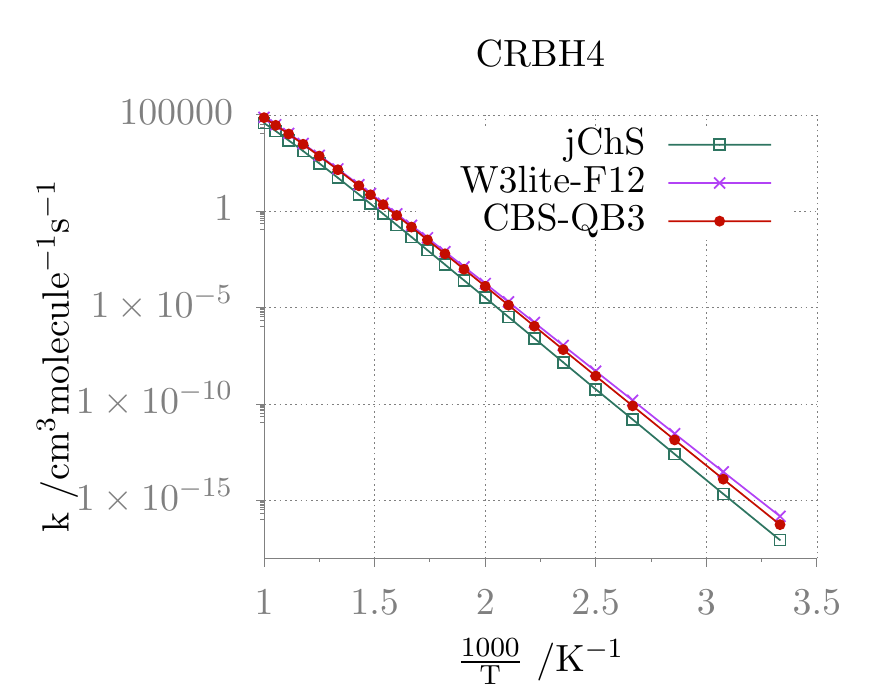}
            \caption[]%
            {{\small CRBH4 $k(T)$}}    
            \label{fig:ratecrbh4}
        \end{subfigure}
        \caption[ Dngty ]
        {Rate constants temperature-dependence plots of the BHPERI1 and CRBH4 reactions from BH14 dataset for a pressure of 1 atm.} 
        \label{fig:rate_bh14}
    \end{figure*}

We next consider the BHPERI1 and CRBH4 reactions discussed in the section on the energy barriers (see Figure \ref{fig:BHPERI1} and \ref{fig:CRBH4}). The rates computed in the 300-1000 K temperature interval by different electronic structure methods are shown in Figure \ref{fig:ratebhperi1} and \ref{fig:ratecrbh4}, whereas the parameters of the corresponding Arrhenius-Kooij fits are collected in Table \ref{tab:bh14_kooij}. Both reactions are characterized by quite high energy barriers and their rates show a clear Arrhenius behaviour. In these circumstances the different electronic structure methods deliver comparable results over the whole temperature range.

\begin{table}[ht]
\centering
\caption{The Arrhenius–Kooij parameters for BHPERI1 and CRBH4 reactions from BH14 dataset.}
\label{tab:bh14_kooij}
\resizebox{\textwidth}{!}{%
\begin{tabular}{@{}lllllll@{}}
\toprule
\multicolumn{1}{c}{}                                  & \multicolumn{3}{c}{BHPERI1}                                                             & \multicolumn{3}{c}{CRBH4}                                                               \\ \midrule
                                                      & \multicolumn{1}{c}{jChS} & \multicolumn{1}{c}{W3lite-F12} & \multicolumn{1}{c}{CBS-QB3} & \multicolumn{1}{c}{jChS} & \multicolumn{1}{c}{W3lite-F12} & \multicolumn{1}{c}{CBS-QB3} \\ \midrule
$A$/\SI{}{\cubic\centi\metre\per\molecule\per\second} & \num{7.43d13}                & \num{8.18d13}                       & \num{1.26d14}                    & \num{1.24d16}                 & \num{2.58d16}                       & \num{1.14d17}                    \\
$n$                                                   & \num{-1.05}               & \num{-1.12}                      & \num{-1.44}                   & \num{-2.81}                & \num{-3.37}                      & \num{-4.02}                   \\
$E$/\SI{}{\kilo\cal\per\mol}                          & \num{34.31}                & \num{34.41}                       & \num{33.53}                    & \num{45.60}                 & \num{44.35}                       & \num{45.83}                    \\
rms                                                   & \num{7.67d-2}                & \num{7.80d-2}                      & \num{1.06d-1}                    & \num{1.05d-1}                 & \num{1.08E-01}                       & \num{1.06E-01}                    \\ \bottomrule
\end{tabular}%
}
\end{table}

%%%%

\begin{figure}
         \centering
         \includegraphics[width=0.6\textwidth]{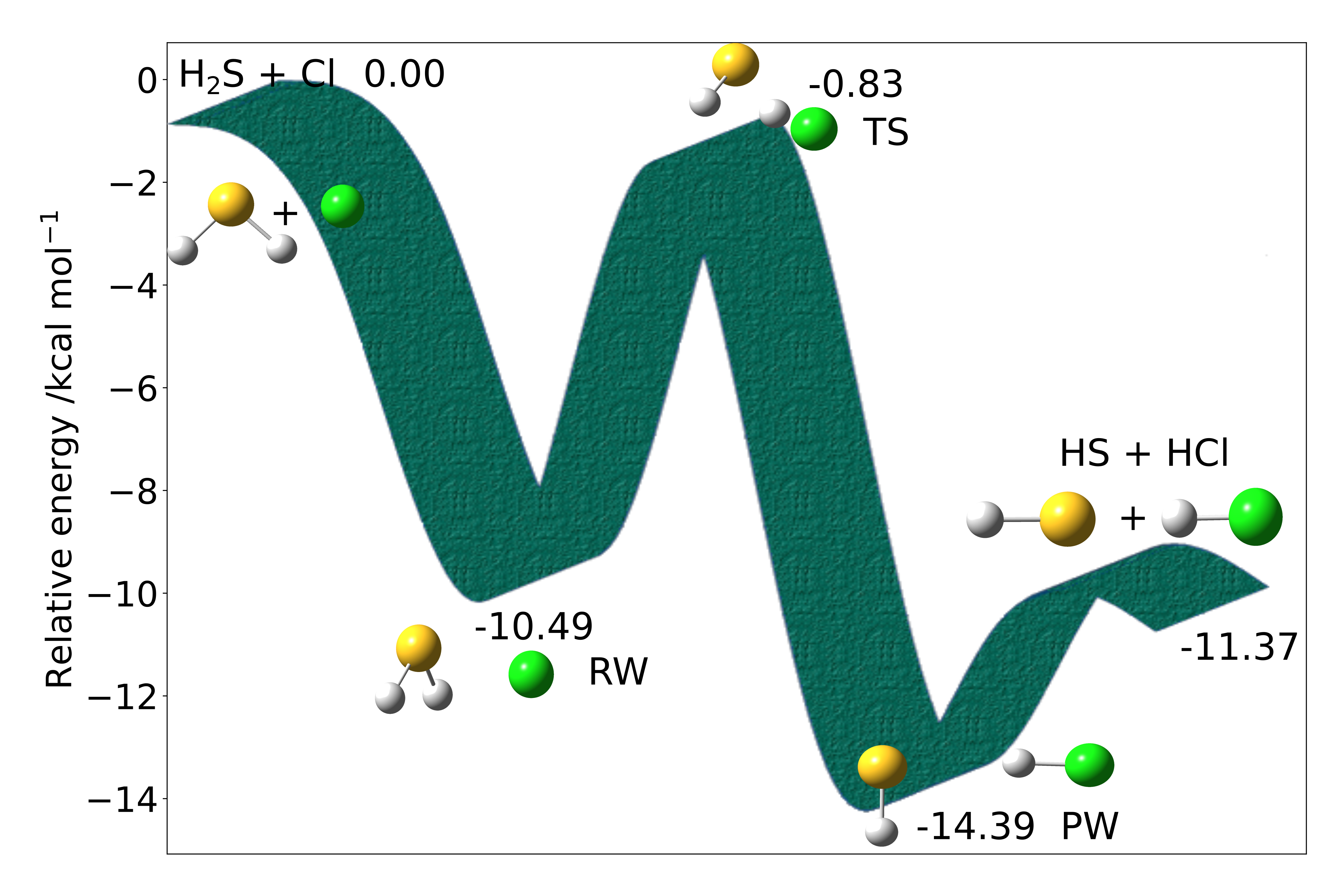}
         \caption{\ce{H2S + Cl}  reaction  mechanism. Electronic energies computed at the jChS level.}
         \label{fig:pesh2s}
\end{figure}

The next example is the reactive potential energy surface for \ce{H2S + Cl} (see Figure \ref{fig:pesh2s}), which involves a van der Waals pre-reactive complex (RW) followed by the transition state TS leading to a product-like van der Waals complex (PW) and then to the products, i.e., \ce{HS + HCl}. Since this reaction has been recently investigated at the CBS-CVH level,\cite{lupi2020} it represents a meaningful test for the jChS model chemistry. Once again, the largest deviation from the reference values for all the stationary points is lower than \SI{0.3}{\kilo\cal\per\mol}, to be compared to errors larger than \SI{1}{\kilo\cal\per\mol} especially for transition states at the CBS-QB3 level. Errors of this magnitude can lead to unreliable rate constants, especially for reactions like this where the dynamical bottleneck is located at the inner transition state, as already pointed out in ref. \citenum{lupi2020}.

\begin{figure}
         \centering
         \includegraphics[width=0.6\textwidth]{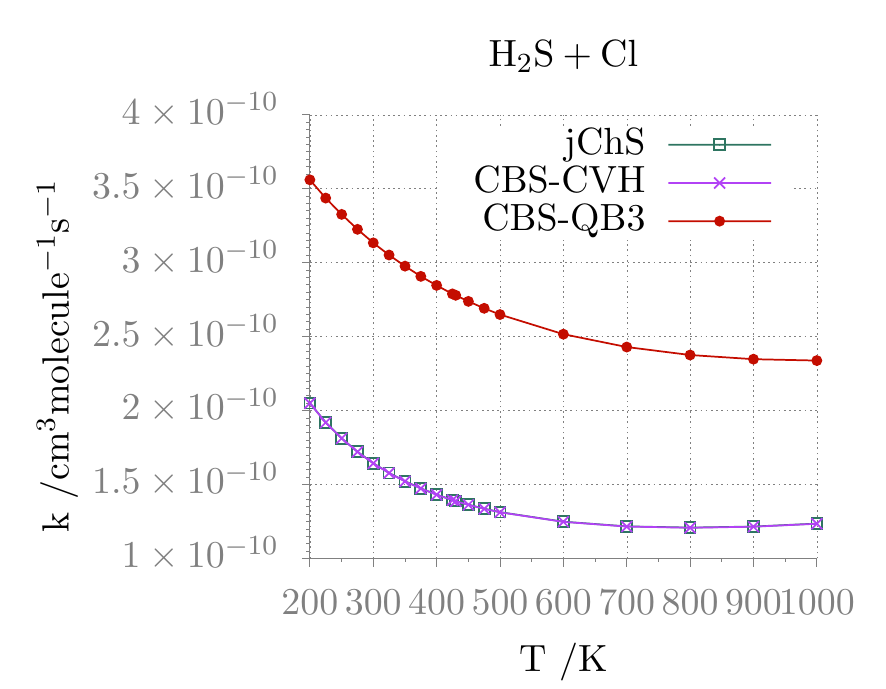}
         \caption{Temperature-dependence plots of the \ce{H2S + Cl} reaction rate constants calculated at various levels of theory for a pressure of 1 atm.}
         \label{fig:rateh2s}
\end{figure}

The reaction rates issuing from jChS computations are compared in Figure \ref{fig:rateh2s} to the CBS-QB3 and CBS-CVH counterparts, whereas the parameters of the corresponding Arrhenius-Kooij fittings (see Equation \ref{eq:kooij}) are collected in Table \ref{tab:my-table}. The root mean square deviations reported in Table \ref{tab:my-table} demonstrate that the data are indeed well fitted by the Arrhenius-Kooij expression with a negative activation energy ($E$) at 0 K. The results issuing from jChS and CBS-CVH computations are virtually indistinguishable, whereas significantly larger rates are obtained at the CBS-QB3 level.

\begin{table}[ht]
\centering
\caption{The Arrhenius–Kooij parameters for the \ce{H2S + Cl} reaction.}
\label{tab:my-table}
\begin{tabular}{@{}lccc@{}}
\toprule
    & jChS   & CBS-CVH & CBS-QB3   \\ \midrule
$A$/\SI{}{\cubic\centi\metre\per\molecule\per\second}   & \num{9.12d-11}  & \num{9.11d-11}  & \num{2.63d-10}  \\
$n$   & \num{7.65d-2}  & \num{7.67d-2}  & \num{-1.60d-1} \\
$E$/\SI{}{\kilo\cal\per\mol}   & \num{-3.42d-1} & \num{-3.42d-1} & \num{-9.94d-2} \\
rms & \num{2.51d-12}  & \num{2.51d-12}  & \num{2.87d-12}  \\ \bottomrule
\end{tabular}
\end{table}

\begin{figure}
         \centering
         \includegraphics[width=\textwidth]{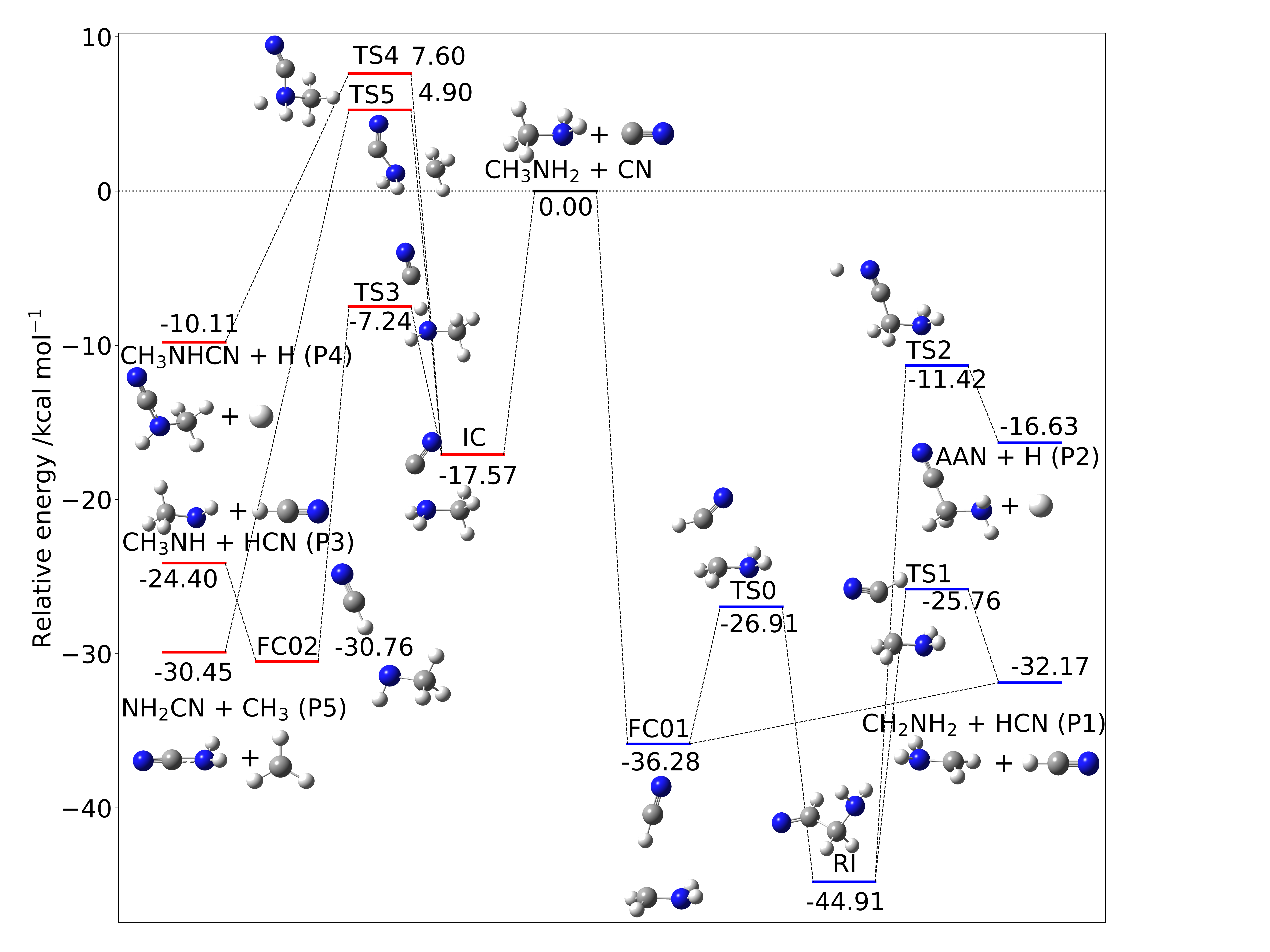}
         \caption{\ce{CH3NH2 + CN} reaction mechanism. The pathway for the attack of CN to the N moiety of methylamine is marked in red while for the abstraction of H from the methyl group by CN in blue. Electronic energies are computed at the jChS level.}
         \label{fig:pesmeth}
\end{figure}

The last example is the quite complex reactive potential energy surface ruling the addition of CN to \ce{CH3NH2} shown in Figure \ref{fig:pesmeth} together with the jChS energies of all the stationary points. The experimental reaction rate at different temperatures \cite{ch3nh2_exp} has been recently well reproduced employing CBS-CVH electronic structure computations within a master equation treatment similar to that employed in the present paper.\cite{puzzarini2020} This system represents, therefore, a challenging test for the jChS model. 

The attack of CN on the nitrogen side of methylamine proceeds via a potential well associated with a pre-reactive complex, \ce{NC\bond{...}NH2CH3} (IC), which evolves in an inner (submerged) transition state (TS3) that, passing through an \ce{NCH\bond{...}NHCH3} intermediate (FC02), forms the \ce{HCN + NHCH3} products (P3). Alternative channels, and in particular that leading to \ce{NH2CN + CH3}, are ruled by non-submerged transition states and are, therefore, closed under the ISM conditions. The attack on the methyl side forms directly the FC01 complex, which, in turn, leads to \ce{HCN + NH2CH2} (P1) without any potential energy barrier. In this case, the alternative two-step mechanism (TS0-RI-TS2-P2) leading to aminoacetonitrile + H is open since it involves only submerged transition states, but it is less favorable.

\begin{figure}[ht]\centering
\subfloat[P1 formation rate constants.]{\label{P1}\includegraphics[width=.45\linewidth]{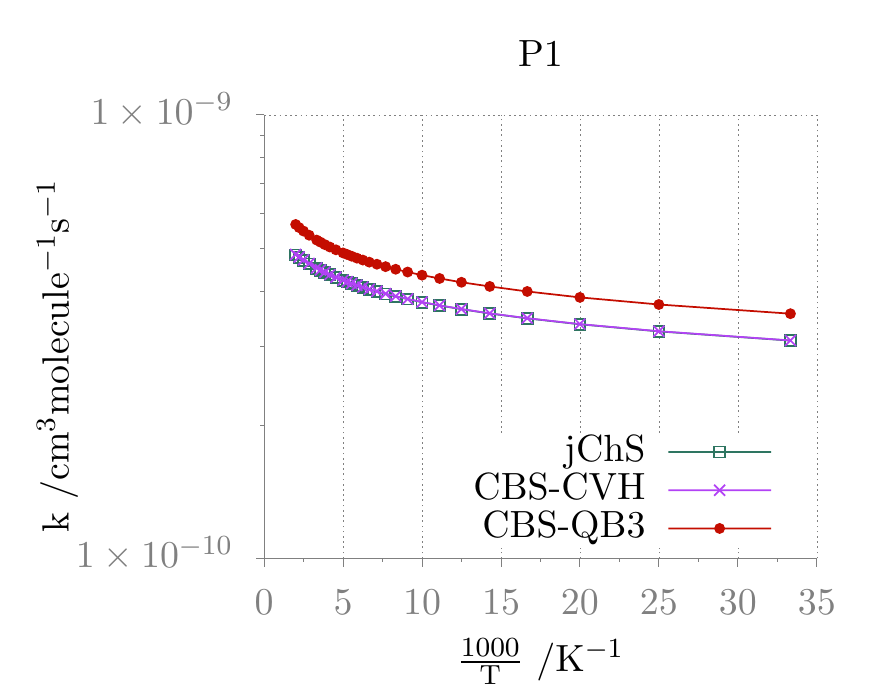}}\hfill
\subfloat[P2 formation rate constants.]{\label{P2}\includegraphics[width=.45\linewidth]{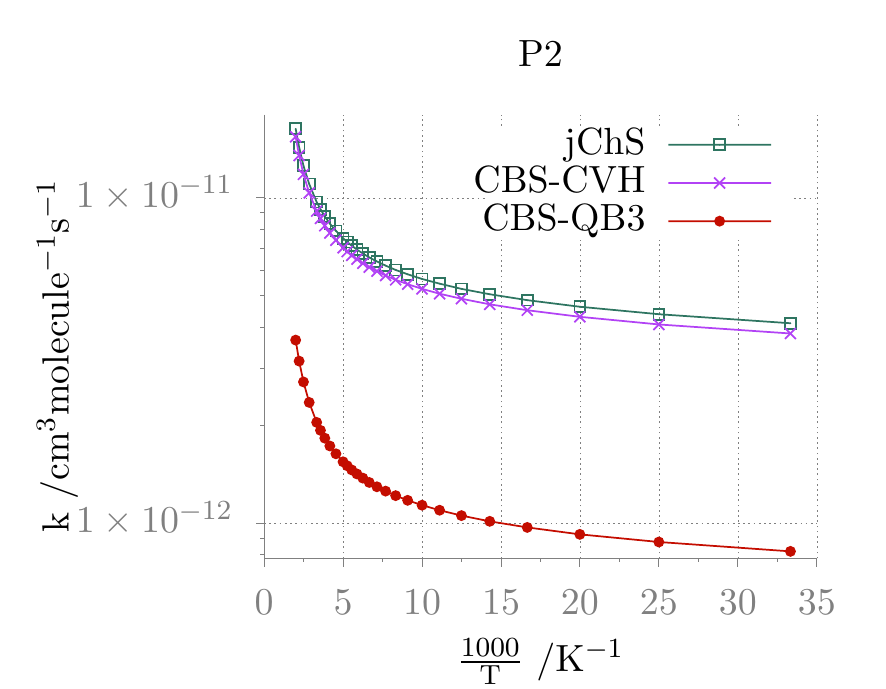}}\par 
\subfloat[P3 formation rate constants.]{\label{P3}\includegraphics[width=.45\linewidth]{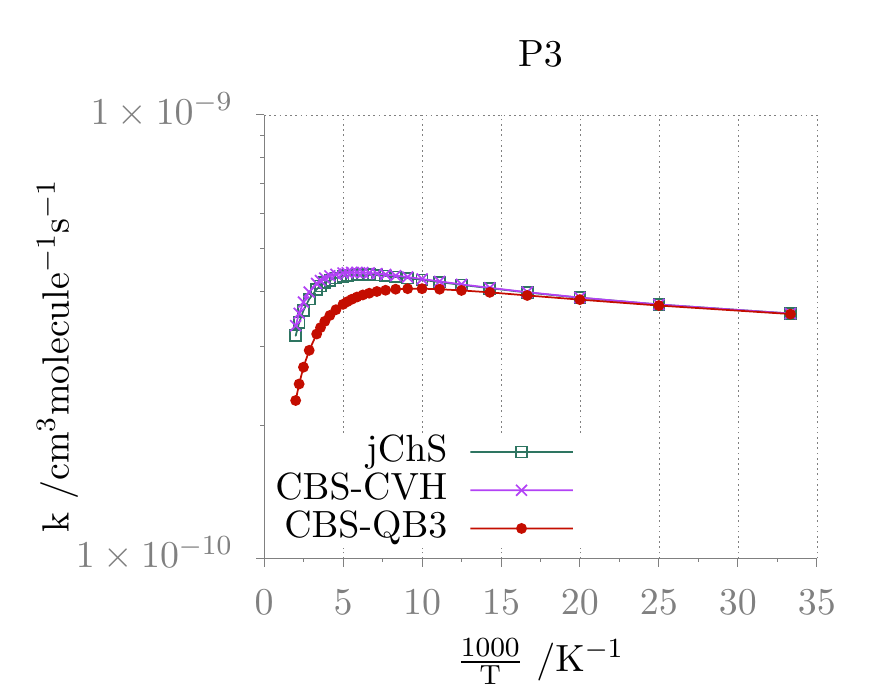}}
\caption{Temperature-dependence plots of the \ce{CH3NH2 + CN} reaction rate constants calculated at various levels of theory for a pressure of 10$^{-8}$ atm.}
\label{fig:rate_met}
\end{figure}

Comparison with the CBS-CVH results of ref.\citenum{puzzarini2020} shows a MAE of \SI{0.26}{\kilo\cal\per\mol} and a maximum deviation of \SI{-0.55}{\kilo\cal\per\mol} for the relative energies of all the stationary points. The errors of the CBS-QB3\cite{montgomery2000} model are again larger than \SI{1}{\kilo\cal\per\mol}, in agreement with the estimates of previous studies. \cite{simmie2015} The reaction rates issuing from jChS computations are compared in Figure \ref{fig:rate_met} with the CBS-QB3 and CBS-CVH counterparts, whereas the parameters of the corresponding Arrhenius-Kooij fits (see Equation \ref{eq:kooij}) are collected in Table \ref{tab:kooijmet}. Noted is that pressure does not influence the reaction rate, as the reactants always proceed to form the products without experiencing significant collisional stabilization in the investigated pressure range (0.001-1 bar).

A curved Arrhenius plot is obtained when the activation energy depends on the temperature and this behavior is captured by the Arrhenius-Kooij equation when this dependence is linear. The root mean square deviations reported in Table \ref{tab:kooijmet} demonstrate that the data for the different products are indeed well fitted by the Arrhenius-Kooij expression. Within this model, $E$ represents the activation energy at 0 K and the activation energy at a generic temperature $T$ is given by $E+n\left(\frac{RT}{300}\right)$. In the present case, the activation energy is positive for P1 and P3,  as a result of both the capture rate constant and the subsequent energy barriers for the unimolecular steps. The value is, instead negative for P2, but in this case the Arrhenius plot is essentially linear. The $n$ parameter (the first derivative of the activation energy with respect to temperature) is positive for the P1 and P2 products, thus reflecting an increase of the activation energy with temperature, while the opposite behavior ($n$ negative) is obtained for P3. Finally, the value of the pre-exponential factor $A$ is typical for this kind of reactions and rules the branching ratio between the different channels. 

\begin{table}[ht]
\centering
\caption{The Arrhenius–Kooij parameters for the \ce{CH3NH2 + CN} reaction.}
\label{tab:kooijmet}
\resizebox{\textwidth}{!}{%
\begin{tabular}{@{}lccccccccc@{}}
\toprule
    & \multicolumn{3}{c}{jChS}       & \multicolumn{3}{c}{CBS-CVH}    & \multicolumn{3}{c}{CBS-QB3}      \\ \midrule
    & P1       & P2        & P3        & P1       & P2        & P3        & P1       & P2        & P3        \\ \cmidrule(l){2-10} 
$A$/\SI{}{\cubic\centi\metre\per\molecule\per\second}   & \num{4.51d-10} & \num{8.95d-12}  & \num{4.28d-10}  & \num{4.52d-10} & \num{8.38d-12}  & \num{4.38d-10}  & \num{5.22d-10} & \num{1.86d-12}  & \num{3.54d-10}  \\
$n$   & \num{1.50d-1} & \num{8.70d-1}  & \num{-2.43d-1} & \num{1.51d-1} & \num{8.84d-1}  & \num{-2.09d-1} & \num{1.63d-1} & \num{9.78d-1}  & \num{-4.45d-1} \\
$E$/\SI{}{\kilo\cal\per\mol}   & \num{2.10d-3} & \num{-8.31d-2} & \num{4.99d-2}  & \num{2.02d-3} & \num{-8.48d-2} & \num{4.63d-2}  & \num{5.28d-4} & \num{-9.72d-2} & \num{6.71d-2}  \\
rms & \num{9.70d-13} & \num{4.54d-13}  & \num{1.75d-11}  & \num{9.40d-13} & \num{4.33d-13}  & \num{1.69d-11}  & \num{3.31d-13} & \num{1.11d-13}  & \num{1.73d-11}  \\ \bottomrule
\end{tabular}%
}
\end{table}

\section{Conclusions}
Astrochemistry and atmospheric chemistry require accurate kinetic data for processes occurring at low to moderate temperatures and involving barrier heights spanning a large range of values. Furthermore, the chemical species involved in these processes can contain more than ten non-hydrogen atoms and non-covalent interactions may often rule the entrance channels. We have, therefore, developed and validated a new general model rooted into the master equation formalism employing ab-initio transition state theory for computing the reaction rates of elementary processes. To this end, we have slightly modified and validated the recently proposed jChS model chemistry with reference to very accurate energetic and kinetic data. The results obtained for a large panel of systems and reaction channels show an average error
within \SI{0.3}{\kilo\cal\per\mol} without the need of any empirical parameter, which allows the evaluation of accurate branching ratios and leads to errors within 20$\%$ for reaction rates. 

The computational bottleneck of the proposed model chemistry is the CCSD(T)/jun-cc-pVTZ step and, in this connection, recent localized treatments of correlation (for example using local pair natural orbitals
\cite{lpno1,lpno2}) will be investigated in order to further increase the dimension of molecular systems amenable to accurate computations with reasonable computer requirements. Furthermore, the performances of the jChS model for other classes of reactions of particular interest for astrochemistry and/or atmospheric chemistry (e.g. those involving ozone and Criegee intermediate) need be investigated in deeper detail. In the same vein, further refinements and validations are surely needed for specific situations (e.g., non negligible static correlation effects or intersystem crossing). However, even taking these caveats into account,  we think that the results reported in the present paper pave the route for the accurate study of chemical processes under widely different temperature and pressure conditions. 
%%%%%%%%%%%%%%%%%%%%%%%%%%%%%%%%%%%%%%%%%%%%%%%%%%%%%%%%%%%%%%%%%%%%%
%% The "Acknowledgement" section can be given in all manuscript
%% classes.  This should be given within the "acknowledgement"
%% environment, which will make the correct section or running title.
%%%%%%%%%%%%%%%%%%%%%%%%%%%%%%%%%%%%%%%%%%%%%%%%%%%%%%%%%%%%%%%%%%%%%
\begin{acknowledgement}
This work has been supported by MIUR (Grant Number 2017A4XRCA), by the Italian Space Agency (ASI; ‘Life in Space’ project, N. 2019-3-U.0) and by Scuola Normale Superiore (SNS18$_-$B$_-$Tasinato). The SMART@SNS Laboratory (\url{http://smart.sns.it}) is acknowledged for providing high-performance computing facilities. The authors thank Dr. Silvia Alessandrini and Prof. Cristina Puzzarini (University of Bologna) for useful discussions. 
\end{acknowledgement}

%%%%%%%%%%%%%%%%%%%%%%%%%%%%%%%%%%%%%%%%%%%%%%%%%%%%%%%%%%%%%%%%%%%%%
%% The same is true for Supporting Information, which should use the
%% suppinfo environment.
%%%%%%%%%%%%%%%%%%%%%%%%%%%%%%%%%%%%%%%%%%%%%%%%%%%%%%%%%%%%%%%%%%%%%
\begin{suppinfo}
The Supporting Information is available free of charge at xxxxxxxx.
Extended Table of results for the HTBH38/08 dataset, cartesian coordinates of all the stationary points optimized in this work and MESS input files.
\end{suppinfo}

%%%%%%%%%%%%%%%%%%%%%%%%%%%%%%%%%%%%%%%%%%%%%%%%%%%%%%%%%%%%%%%%%%%%%
%% The appropriate \bibliography command should be placed here.
%% Notice that the class file automatically sets \bibliographystyle
%% and also names the section correctly.
%%%%%%%%%%%%%%%%%%%%%%%%%%%%%%%%%%%%%%%%%%%%%%%%%%%%%%%%%%%%%%%%%%%%%
\bibliography{achemso-demo}

\end{document}